\documentclass[twocolumn]{aastex631} 
\usepackage[utf8]{inputenc}
\usepackage{longtable}
\usepackage{amsmath}
\usepackage{graphicx}

\usepackage[bottom]{footmisc}

\defcitealias{huangrui2023APJS}{Paper I}

\shortauthors{R. Huang et al.}
\shorttitle{XMM-Newton Legacy Survey of M31}

\begin{document}

\title{An XMM-{New}ton View of the {AN}dromeda {G}alaxy as {E}xplored in a {L}egacy {S}urvey (New-ANGELS) II: Luminosity Function of X-ray Sources}

\author[0000-0001-7900-4204]{Rui Huang}
\affiliation{Department of Astronomy, Tsinghua University, Beijing 100084, China}
\affiliation{Department of Astronomy, University of Michigan, 311 West Hall, 1085 S. University Ave, Ann Arbor, MI, 48109-1107, U.S.A.}

\author[0000-0001-6239-3821]{Jiang-Tao Li}
\affiliation{Purple Mountain Observatory, Chinese Academy of Sciences, 10 Yuanhua Road, Nanjing 210023, China}

\author[0000-0002-6324-5772]{Wei Cui}
\affiliation{Department of Astronomy, Tsinghua University, Beijing 100084, China}

\author[0000-0002-2941-646X]{Zhijie Qu}
\affiliation{Department of Astronomy, Tsinghua University, Beijing 100084, China}
\affiliation{Department of Astronomy \& Astrophysics, The University of Chicago, Chicago, IL 60637, U.S.A}

\author[0000-0001-6276-9526]{Joel N. Bregman}
\affiliation{Department of Astronomy, University of Michigan, 311 West Hall, 1085 S. University Ave, Ann Arbor, MI, 48109-1107, U.S.A.}



\author[0000-0002-0584-8145]{Xiang-Dong Li}
\affiliation{School of Astronomy and Space Science, Nanjing University, Nanjing 210023, People's Republic of China}
\affiliation{Key Laboratory of Modern Astronomy and Astrophysics, Nanjing University, Ministry of Education, Nanjing 210023, People's Republic of China}

\author[0000-0003-0293-3608]{Gabriele Ponti}
\affiliation{INAF-Osservatorio Astronomico di Brera, Via E. Bianchi 46, 23807 Merate (LC), Italy}
\affiliation{Max-Planck-Institut für Extraterrestrische Physik, Giessenbachstrasse, 85748 Garching, Germany}

\author[0000-0002-9279-4041]{Q. Daniel Wang}
\affiliation{Department of Astronomy, University of Massachusetts, Amherst, MA 01003, U.S.A.}


\correspondingauthor{Jiang-Tao Li}
\email{pandataotao@gmail.com}

\begin{abstract}

As part of the New-ANGELS program, we systematically investigate the X-ray luminosity functions (XLFs) of 4506 X-ray sources projected within a radius of 2.5 deg centering on M31. We construct XLFs for different regions in the disk and halo of M31, accounting for the incompleteness with an effective sensitivity map. Assuming that the halo regions contain (mostly) foreground stars and background active galactic nuclei, they are taken as "background" for deriving the XLFs of the sources in the disk. Through modeling XLFs, we decompose the X-ray sources into distinct populations for each region. We find that low-mass X-ray binaries are the dominant X-ray population throughout the disk of M31.
The XLFs of M31 reveal a consistently lower integrated LMXB luminosity per stellar mass ($\alpha_\mathrm{LMXB}$) compared to other galaxies, likely due to M31's prolonged period of {{quiescent}} star formation. Variations in the XLF shape and $\alpha_\mathrm{LMXB}$ across different regions of M31 suggest that the relationship between integrated luminosity and stellar mass may vary within the galaxy. Additionally, the relatively low integrated luminosity observed in the inner-arm region provides crucial evidence for a rapid fading of M31's LMXBs around 1~Gyr, a finding consistent with recent observations of other nearby galaxies.

\end{abstract}


\section{Introduction} \label{sec:Intro}

{Stellar X-ray sources are a key component of galactic X-ray emission, distinct from the central AGN. These sources primarily include} 
low-mass X-ray binaries (LMXBs; \citealt{gilfanovLowmassXrayBinaries2004,Lehmer2019}), high-mass X-ray binaries (HMXBs; \citealt{Lehmer2019,Mineo2012HMXB}), cataclysmic variables {(CV; \citealt{Mukai2017,Pretorius2012}}), and coronally active binaries (AB; \citealt{Sazonov2006}).
These X-ray sources not only offer direct insights into the evolved binary components of different stellar populations but also can be linked to 
various galaxy properties (e.g., star formation rate, stellar mass, and metallicity).

The properties of LMXB and HMXB populations are physically correlated with the galaxy's star formation history.
The X-ray-emitting timescale of LMXBs is long ($\rm \gtrsim 1$ Gyr, due to the long lifetime of their low-mass donor stars). 
The total number or luminosity of LMXBs therefore depends on the galaxy's stellar mass \citep{Lehmer2019}.
In contrast, HMXBs stay X-ray bright within at most 100 million years after the formation of their high-mass companions, so their abundance is tightly correlated with the galaxy's very recent star formation rate (SFR;  \citealt{Sunyaev1978,Grimm2003,Mineo2012HMXB}).
In addition to the star formation history traced by stellar mass and SFR, the formation and evolution of the LMXBs and HMXBs may also be influenced by other galaxy properties, such as the forming environment \citep{Zhang2011}, stellar age \citep{Zhang2012}, and metallicity \citep{Lehmer2021metal}. 

The X-ray luminosity functions (XLFs) of stellar sources in nearby galaxies are a powerful tool to study X-ray source populations, therefore providing valuable insight into galaxy evolution \citep{fabbianoPopulationsXRaySources2006,Gilfanov2022}. 
Previous studies typically focused on dwarf galaxies or starburst galaxies because of their abundant LMXB or HMXB populations.
XLFs are constructed over entire galaxies or even combined dozens of galaxies to address the low statistics of detected X-ray sources (e.g., \citealt{Grimm2003, gilfanovLowmassXrayBinaries2004, Zhang2012, Mineo2012HMXB}).
The connection with galaxy properties is therefore limited to galactic-scale characteristics.

Spatially resolved studies have become available only in recent years (e.g., \citealt{Antoniou2010,Lehmer2014, Lehmer2019}). 
M31 is unique in exploring the spatially resolved XLF in single galaxies because of its proximity, 
abundant detected sources, and extensive multiwavelength observations.
With a distance of 761 kpc \citep{liSub2DistanceM312021}, M31 promises the construction of the XLF for various X-ray populations down to a limiting luminosity of $10^{35}\rm~erg~s^{-1}$, which is orders of magnitude more sensitive than the limits in other local galaxies of $10^{36}-10^{38}\rm~erg~s^{-1}$ (e.g., \citealt{gilfanovLowmassXrayBinaries2004}, \citealt{Lehmer2019}).
Furthermore, the small relative uncertainty of M31 distance ($\lesssim 1.5\%$) enables more accurate luminosity measurements, which is in contrast to many Galactic sources (e.g., \citealt{revnivtsevOriginGalacticRidge2006}). 
M31's higher stellar mass results in a greater number of stellar X-ray sources compared to most other external galaxies (e.g., \citealt{gilfanovLowmassXrayBinaries2004},\citealt{vossLuminosityFunctionXray2006}). 
The tremendous multi-wavelength studies (i.e.,  \citealt{ LGGS_Massey2006,Williams2017PHAT,Jarrett2019}) provide us with in-depth insights into the various properties of M31, expanding our related exploration.

The XLF studies in M31 have spanned decades.
{These investigations primarily focus on the bulge due to its high source density. 
The bulge of M31 has an effective radius of $\sim4.5\arcmin$ 
(equivalent to $1.0\pm0.2$~kpc; \citealt{Courteau2011}). 
A typical central region for X-ray studies, often extending to an $8\arcmin$ 
radius, thus covers most of the bulge component (see also Figure~9 in \citealt{Courteau2011}). 
Both can be largely encompassed by a single \emph{XMM-Newton} 
field of view (radius $\sim15\arcmin$), and are also well covered by \emph{Chandra} 
($17\arcmin \times 17\arcmin$ FOV).}
There is ongoing debate regarding the galactocentric distance dependence of the XLF in the central region, where LMXBs dominate.
In the initial \emph{Chandra} catalog of M31's central region ($17^{\prime} \times 17^{\prime}$), presented by \cite{kongXRayPointSources2002}, distinct XLF patterns are revealed in different regions (inner bulge: $2^{\prime} \times 2^{\prime}$, outer bulge: $8^{\prime} \times 8^{\prime}$ with inner bulge excluded, and disc: $17^{\prime} \times 17^{\prime}$ with central $8^{\prime} \times 8^{\prime}$ excluded). 
The inner bulge displays a break at $10^{36}\rm~erg~s^{-1}$, shifting towards higher luminosity from the inner bulge to the disc. 
In contrast, \cite{Trudolyubov2002} utilize \emph{XMM–Newton} data and identify similar XLF slopes for both the disc and bulge, considering a 15 arcmin radius for the bulge.
Another study by \cite{vossStudyPopulationLMXBs2007} finds that within a 12 arcmin radius, the luminosity functions of field
LMXBs are distance-independent.
Moreover, apart from the observed spatial correlation between field LMXBs and the K-band light, \cite{vossStudyPopulationLMXBs2007} observes that the spatial distribution of LMXBs within one arcmin of M31's nucleus is directly proportional to the square of the K-band light profile, indicating their dynamical forming origin.
\cite{Zhang2011} further confirms that the XLF of M31's nucleus sources resembles that of sources formed in globular cluster, which is also in high-density environments compared to in the field. More recent studies have continued to refine our understanding. Early XMM-Newton observations of the disk provided qualitative descriptions of the XLF shape \citep{greeningXraySpectralSurvey2009}, while recent NuSTAR observations have begun to probe the hard X-ray ($E>10$ keV) emission from both the disk \citep{Moon2024} and as part of larger galaxy samples that establish universal scaling relations \citep{Vulic2018}. For a direct comparison of LMXB populations, the most crucial benchmark remains the well-characterized XLF of our own Milky Way (e.g., \citealt{Voss2010}).

Although the studies mentioned above are significant and thorough, a comprehensive and dedicated population study of the entire M31 is still lacking. Previous studies consistently remain confined to specific regions (i.e., bulge, part of disk). Even the previous M31 survey conducted by \cite{stieleDeepXMMNewtonSurvey2011a}, which covers the entire optical disk of M31, does not extend its analysis to the XLF. Additionally, the over-subtraction of AGN population \citep{luoCHANDRADEEPFIELDSOUTH2016}, as mentioned in \cite{vulicXRaysBewareDeepest2016}, complicates the XLF studies within the disk.

An \emph{XMM-Newton} View of the ANdromeda Galaxy as Explored in a Legacy Survey (New-ANGELS) is a legacy X-ray survey of the M31 disk and bulge, as well as its inner halo around the disk. 
This initiative stems from the \emph{XMM-Newton} AO-16 large program (PI Jiang-Tao Li), focusing on surveying the M31 halo within approximately 30 kpc from the center.
In the first paper \citep[][hereafter, \citetalias{huangrui2023APJS}]{huangrui2023APJS}, 
we detect a remarkable total of 4506 sources over 7.2 square degrees around M31. This unprecedented catalog not only facilitates a comprehensive XLF study across various regions of M31 but also provides a rare and valuable background region for these investigations. 
In this paper, we extract the XLFs of various X-ray populations in different regions (i.e., center, disk, north, and south), and investigate their connections with stellar properties.
This paper is organized as follows:
In \S\ref{sec:incompleteness-correction}, we detail the methodology for constructing the XLF and address incompleteness issues. The results of the XLF analysis are presented in \S\ref{sec:Result}, followed by in-depth discussions in \S\ref{sec:Discussion}. 

\section{Construction of X-ray Luminosity Functions}
\label{sec:incompleteness-correction}

Based on the X-ray source catalog constructed in \citetalias{huangrui2023APJS}, we herein describe how to extract XLFs of these sources in different regions, and how to correct for some biases. 
The XLF represents the total number of sources above a specific luminosity threshold, presented as a function of luminosity. We calculate the XLF in distinct energy bands: 0.2-0.5 keV, 0.5-1.0 keV, 1.0-2.0 keV, 2.0-4.5 keV, and 4.5-12.0 keV, following the energy band configurations outlined in \citetalias{huangrui2023APJS}.

\subsection{Incompleteness of XLFs}
The incompleteness correction is crucial for calculating XLFs, because faint sources can only be detected in particular regions.
First, exposure times and background levels vary across different observations in our survey, so faint sources are preferentially detected in deep observations.
Then, the vignetting effect and the off-axis point spread function (PSF) make the detection limit uneven in individual observations.
Consequently, faint sources are reliably detectable near the optical axis, with detectability decreasing as they move to larger off-axis angles. 

For {contaminating} sources (i.e., AGN, foregroundstars) that are uniformly distributed in our field, the incompleteness of the X-ray luminosity function can be characterized by the sky coverage function, which quantifies the sky area at certain X-ray flux-limit levels. This information can be derived from simulations or sensitivity maps.
However, simulating source detection presents computational challenges, especially when dealing with multiple overlapping observations, as in the case of the M31 field. For example, source detection on half of the M31 disk as in \citetalias{huangrui2023APJS} needs $\sim100$ CPU hours. 
{Furthermore, to avoid source confusion, we only insert 10-20 new sources per observation, which means thousands of simulations would be required to generate a statistically sufficient number of sources for accurate incompleteness assessment. Such computational demands are prohibitive with our available resources.}

To tackle this challenge, we choose to create analytical sensitivity maps, which are more efficient and convenient, allowing us to correct the incompleteness after generating sensitivity maps with several hours in any designed regions.
However, the sensitivity map provided by SAS task \texttt{esensmap}\footnote{\url{https://xmm-tools.cosmos.esa.int/external/sas/current/doc/esensmap}} is tailored for sliding box source detection in a single observation rather than maximum likelihood PSF fitting across multiple overlapping observations, as employed in our catalog (\citetalias{huangrui2023APJS}).
To overcome this limitation, we develop a method to create sensitivity maps specifically for source detection in overlapping observations using the SAS task \texttt{edetect\_stack}
\footnote{\url{https://xmm-tools.cosmos.esa.int/external/sas/current/doc/edetect\_stack}}. 
Importantly, it is fully compatible with the standalone task \texttt{emldetect}
\footnote{\url{https://xmm-tools.cosmos.esa.int/external/sas/current/doc/emldetect}}
, which conducts \emph{XMM-Newton} EPIC maximum likelihood multi-source point spread function fitting and serves as a vital component of \texttt{edetect\_stack}.
While being designed for \texttt{edetect\_stack}, the underlying concept remains applicable to other source detection algorithms based on C-statistics.
The methodology specifics are outlined in Appendix~\ref{sec:sensitivity_map}. We generate the sensitivity map corresponding to the source detection likelihood (\texttt{EP\_i\_DET\_ML}, see Equation~\ref{eq.L}) in each band (i=1, 2, 3, 4, and 5) above 6 for reliable detection. {This detection likelihood threshold of 6 corresponds to a false detection rate of $<0.5\%$ \citep{Ni2021}, striking a good balance between completeness and reliability in our source detection.}

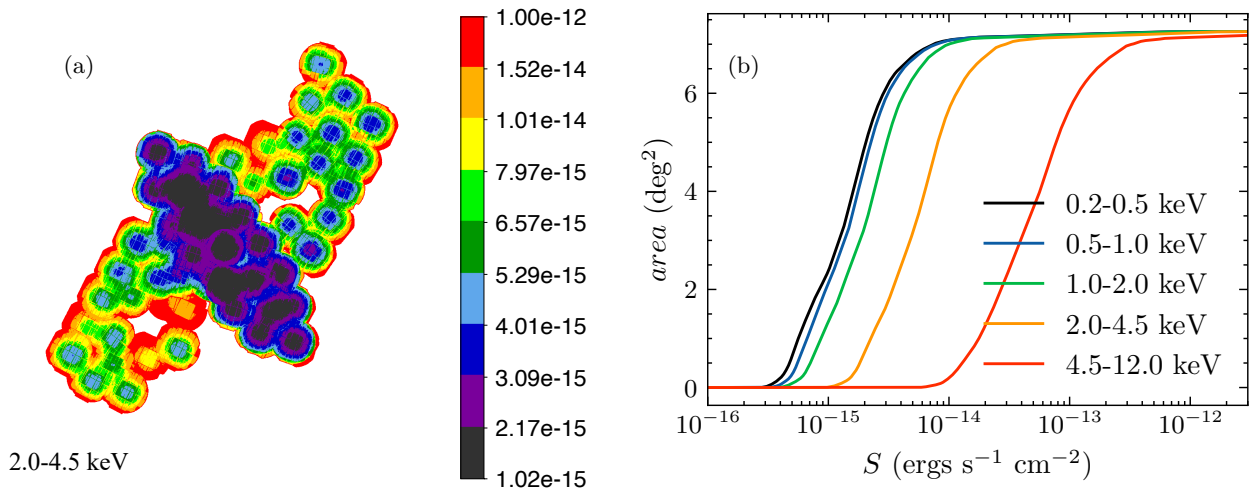
\begin{figure*}
\begin{center}
\includegraphics[width=2.0 \columnwidth]{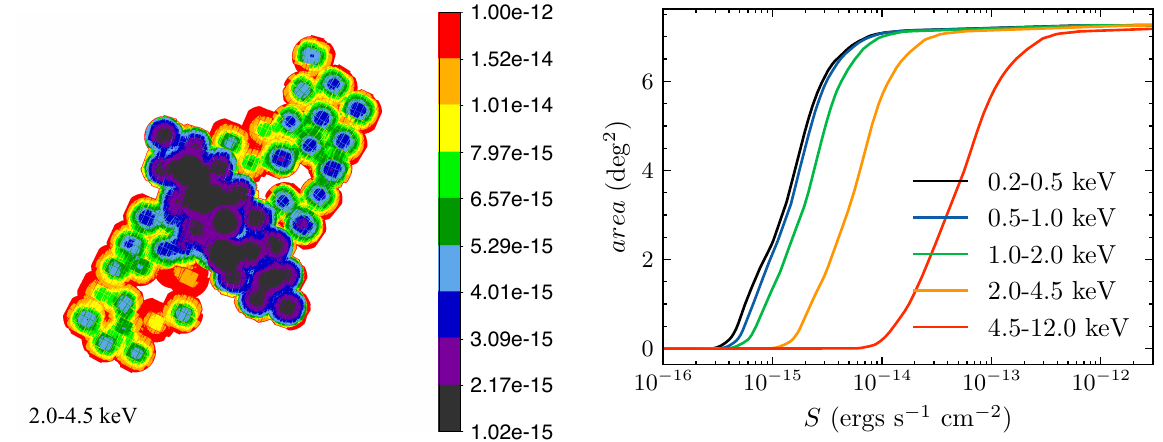}
\begin{picture}(0,0)
\put(-455,160){\makebox(0,0)[l]{{\color{black}(a)}}}
\end{picture}
\begin{picture}(0,0)
\put(-210,160){\makebox(0,0)[l]{{\color{black}(b)}}}
\end{picture}
\end{center}
\caption{(a): The sensitivity map in the energy range of 2.0-4.5 keV; (b): The sky coverage function generated from the sensitivity maps in the  energy ranges of 0.2-0.5 keV, 0.5-1.0 keV, 1.0-2.0 keV, 2.0-4.5 keV, and 4.5-12.0 keV.} \label{pic:sensitivity_map}
\end{figure*}

Using the sensitivity map, we derive the sky coverage function, presenting the sky area where the sensitivity is below the X-ray flux $S$.
The sensitivity map and the corresponding sky coverage function between 2.0-4.5 keV are presented in Fig.~\ref{pic:sensitivity_map}.

Furthermore, if the population (i.e., LMXBs, HMXBs) is not uniformly distributed over our FoV, we need to consider their intrinsic spatial distribution. For the old population (i.e., LMXB) which is related with stellar mass, we thus derive the stellar mass function based on the sensitivity map and stellar mass map. The stellar mass function is the integrated stellar mass as a function of source X-ray flux over the region where the X-ray flux is above the sensitivity.  For the young population (i.e., HMXB) which is related with SFR,
we define the SFR function in the same approach with sensitivity map and SFR map. The stellar mass map and SFR map are generated with \emph{WISE} W1 and W4 images, details can be seen in Appendix~\ref{sec:stellarmass_SFR}.

After deriving the sky coverage, mass, and SFR functions, we can forward-model the X-ray flux distribution of our sources ($\mathrm{d}N(S)$) by convolving the model of uniformly distributed population with the sky coverage function, and the LMXB and HMXB components with the stellar mass function and SFR function, respectively. 
This approach is taken so that we can use C-statistic \citep{Kaastra2017} to quantify the goodness of the fit, to account for statistics of low counts at both the high and low ends of the differential XLF.

\subsection{Extracting Differential and Cumulative XLFs}

Before generating the XLFs, we first filter out sources with X-ray fluxes below the sensitivity at their respective positions, as determined by the sensitivity map. 
This alleviates the Eddington bias, which is the selection bias that intrinsic faint sources around the detection limit could be detected brighter due to the detection uncertainty.
Therefore, the number of included sources in the XLF analysis 
is less than the total detected sources. This is because we specifically require a source detection likelihood above 6 in the specific band used, instead of considering the combined detection likelihood above 6 from all five bands as our catalog required (\citetalias{huangrui2023APJS}). Then, we can construct the differential XLFs $\mathrm{d}\mathcal{N}(S)$ for modeling, which represents the number of sources per unit flux interval as a function of X-ray source flux. 
For the illustration purpose, we also compute the pseudo-incompleteness-corrected cumulative source number density ($N(>S)$) by applying the filtered sources and the sky coverage function to the formula \citep{cappellutiXMMNewtonWide2007}:
\begin{equation}
 N(>S) = \sum_{i=1}^{N_\mathrm{s}(>S)}{\frac{1}{\Omega_i}}  
 \label{eq.XLF_N}
\end{equation}
where $N_\mathrm{s}(>S)$ is the total number of sources in the field with X-ray fluxes above $S$,
and \(\Omega_i\) represents the sky coverage associated with the X-ray flux of the \(i\)-th source, obtainable from the sky coverage function (Fig.~\ref{pic:sensitivity_map}b). This calculation treats all source populations uniformly. While suitable for illustration, this approach may introduce minor visual artifacts in the plotted shapes of components with non-uniform spatial distributions (e.g., LMXBs, HMXBs).
{To ensure robust statistics, the faint-flux limit ($S_\mathrm{{min}}$) for our X-ray Luminosity Functions (XLFs) was set based on the sky coverage function at corresponding regions. We defined $S_\mathrm{{min}}$ as the flux level at which sources can be detected in at least 10\% of our desired survey region. Consequently, our XLFs are constructed for fluxes $S\geq S_\mathrm{min}$, thereby excluding fainter sources detectable only in smaller, highly sensitive regions where statistics would be less reliable}.
Note that the correction here assumes uniform spatial distribution and does not consider the stellar mass function and the SFR function, as the contributions of LMXBs and HMXBs are not yet determined. Even so, the current correction still allows the comparison of the excess of M31 sources over the background sources (i.e., AGN). The further modeling with the stellar mass function and the SFR function is in \S\ref{sec:Discussion}.

\subsection{Notes on bright sources}
\label{subsec:note_on_bright_sources}
Additionally, it is important to acknowledge the presence of spurious detections around bright sources in the catalog presented in \citetalias{huangrui2023APJS}. The PSF fitting residuals from the bright sources are large enough to trigger the search for an additional source. Consequently, for 28 sources with detection likelihoods (\textit{EP\_DET\_ML}, see \citealt{huangrui2023APJS}) above $2\times 10^{4}$ outside the inner 8 arcmin region of M31, we exclude a total of 59 faint sources around them with detection likelihoods (\textit{EP\_DET\_ML}) that are just 1\% of the central bright sources. 

Moreover, the presence of several bright sources can significantly change the bright end shape of the XLF due to the low statistic. In order to concentrate on sources within M31, we exclude identified background or foreground sources that are located outside the inner region of 8 arcmin of M31 and brighter than $S_\mathrm{2.0-4.5~keV}=2\times10^{-13}~\rm erg~s^{-1}~cm^{-2}$ (except Fig.~\ref{pic:LFv4} and \S\ref{subsec:background}). Among these exclusions, six are AGN, one is a galaxy, and the last one is a galaxy cluster. 
The chosen X-ray flux threshold is determined based on the X-ray flux limit of our XLF in the north and south regions (see Fig.~\ref{pic:LFv4}), ensuring that the completeness of faint sources is preserved while mitigating biases introduced by several bright sources. 
The corresponding models for background and foreground sources are thus cut at this limit to match the treatment.

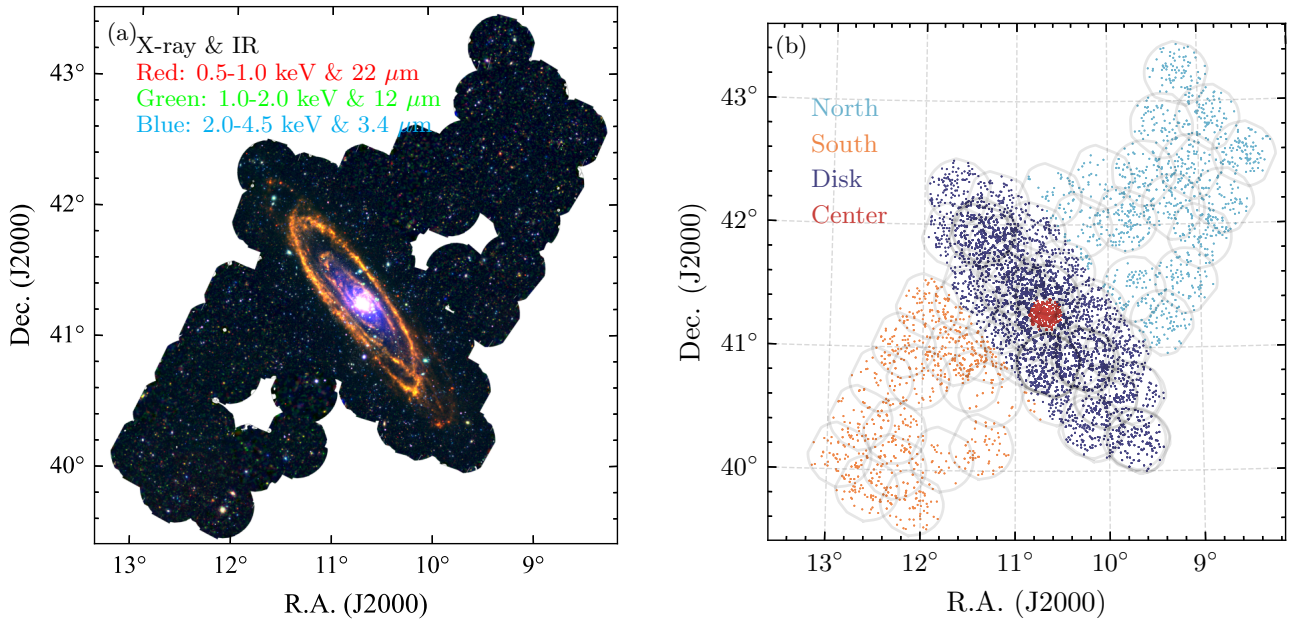
\begin{figure*}
    \centering
    \includegraphics[width=0.98 \columnwidth]{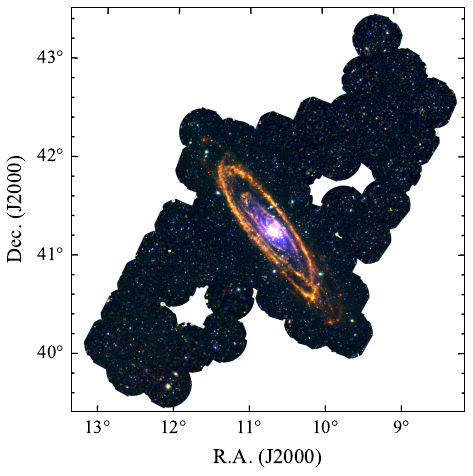}
        \begin{picture}(0,0)
            \put(-200,225){\makebox(0,0)[l]{{\color{black}(a)}}}
        \end{picture}
        \begin{picture}(0,0)
            \put(-195,220){\makebox(0,0)[l]{{\color{black}X-ray \& IR}}}
            \put(-195,210){\makebox(0,0)[l]{{\color{red}Red: 0.5-1.0 keV \& 22 $\rm \mu m$}}}
            \put(-195,200){\makebox(0,0)[l]{{\color{green}Green: 1.0-2.0 keV \& 12 $\rm \mu m$}}}
            \put(-195,190){\makebox(0,0)[l]{{\color{cyan}Blue: 2.0-4.5 keV \& 3.4 $\rm \mu m$}}}
        \end{picture}
    \includegraphics[width=0.98\columnwidth]{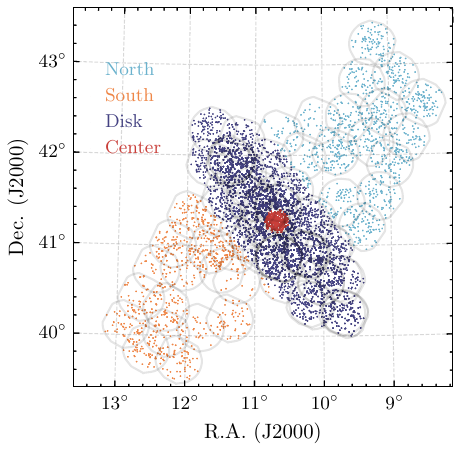}
    \begin{picture}(0,0)
        \put(-200,220){\makebox(0,0)[l]{{\color{black}(b)}}}
    \end{picture}
\caption{(a): A false-color composite image of the M31 survey area, combining X-ray data from \emph{XMM-Newton} and infrared data from \emph{WISE}; 
(b): The observation coverage of M31. The covered area is divided into four regions, as indicated in different colors. The central region is defined as within a radius of $8^{\prime}$ from galactic center, the disk region as outside the central region and along galactic disk, with a height of less than $30^{\prime}$ from galactic center along the minor axis. The north and south regions are the left region in the halo which are $30^{\prime}$ away from the major axis of the galaxy.} \label{pic:region_definition}
\end{figure*}

\section{Results}\label{sec:Result}

\begin{figure*}
\begin{center}
\includegraphics[width=2.1 \columnwidth]{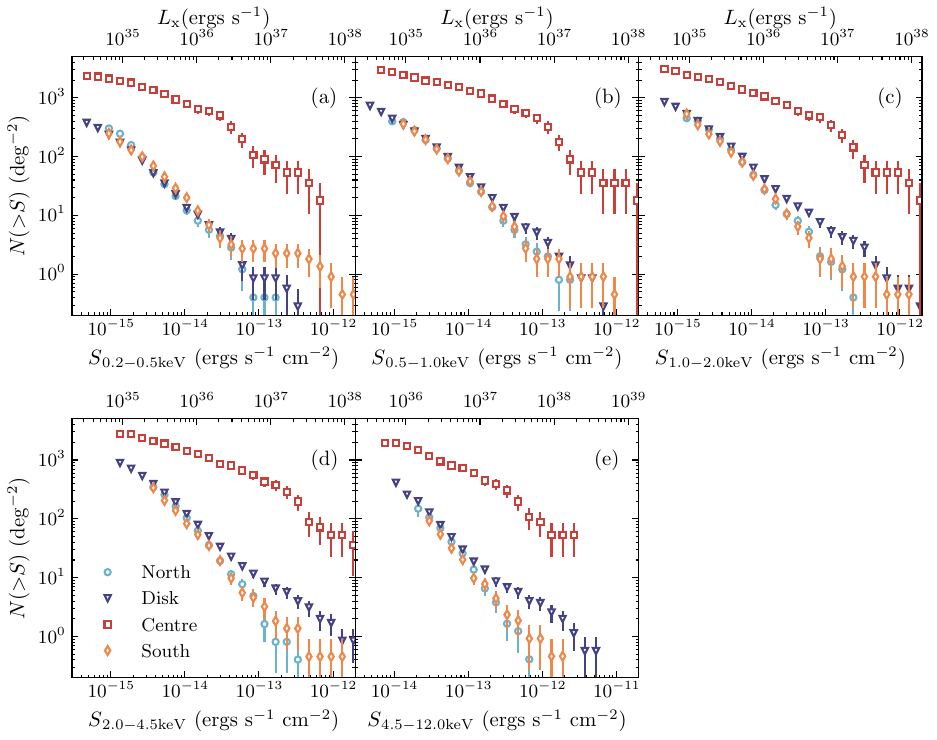}
\end{center}
\caption{X-ray luminosity functions, for the energy ranges of (a) 0.2-0.5 keV, (b) 0.5-1.0 keV, (c) 1.0-2.0 keV, (d) 2.0-4.5 keV, and (e) 4.5-12.0 keV. For comparison, the results are shown for different regions separately.} \label{pic:LFv4}
\end{figure*}

In this section, we construct and analyze XLFs for selected areas in and around M31 (as shown in Fig.~\ref{pic:region_definition}).
The center region is defined as the region within a radius of $8^{\prime}$ from galactic center of M31. The disk region is defined as the area outside the center region and along the galactic disk, with a height of less than 30 arcmin along the minor axis. The north and south regions are located in the halo.
This will enable us to compare the X-ray source population in different areas and maintain roughly uniform exposure depths.
In Fig.~\ref{pic:LFv4}, we present the XLFs corrected for incompleteness in the four regions within the energy ranges of 0.2-0.5 keV, 0.5-1.0 keV, 1.0-2.0 keV, 2.0-4.5 keV, and 4.5-12.0 keV.
The XLFs roughly follow a power law or a broken power law. The XLFs in the north and south regions are quite alike, except at the bright end. 
{The XLF in the disk has a higher normalization than those in the north and south,}
indicating the contribution of the M31 sources. The XLF in the center region has the largest number density and a notable broken structure.

\begin{figure*}
\begin{center}
\includegraphics[width=0.95 \columnwidth]{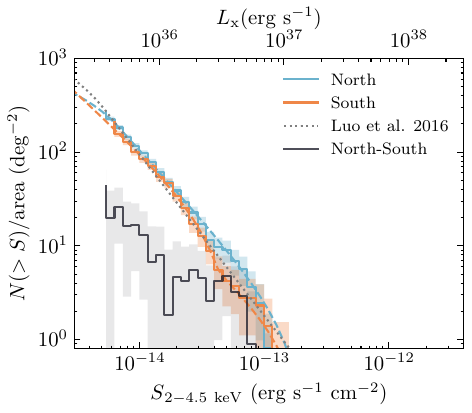}
\includegraphics[width=0.95 \columnwidth]{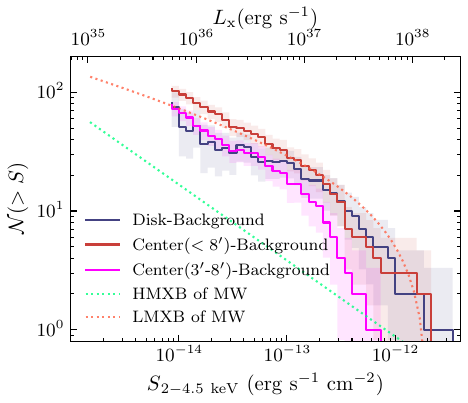}
\end{center}
\caption{X-ray luminosity functions, with the best-fit models (in dashed lines) between 2.0 {and} 4.5 keV. Left: the south (in orange) and north regions (in blue). For comparison, a AGN model (\citealt{luoCHANDRADEEPFIELDSOUTH2016}) is shown in grey dotted line. The black line represents the difference between the XLF of the north and that of the south, obtained by subtraction. Right: XLF of the disk region, central $8^{\prime}$ and the central $3^{\prime}-8^{\prime}$ region, all with the contribution of the sources in the background region (north and south combined) subtracted. Also shown in the latter panels are the XLFs of HMXB (in green dotted line) and LMXB (in red dotted line) in the Milky Way (\citealt{2002A&A...391..923G}), for comparison.} \label{pic:LF_model}
\end{figure*}

\begin{figure*}
    \centering
    \begin{minipage}{0.32 \textwidth}
        \centering
        \includegraphics[width=\linewidth]{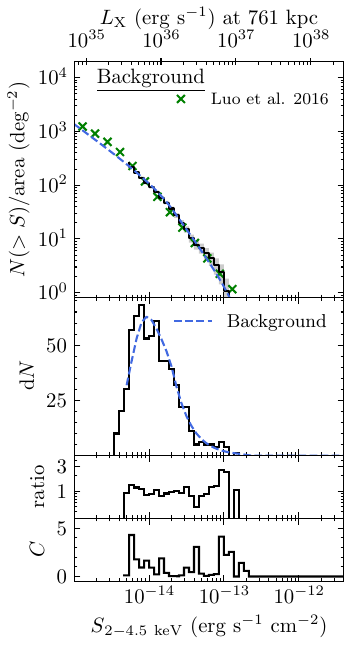}
    \end{minipage}%
    \begin{minipage}{0.32 \textwidth}
        \centering
        \includegraphics[width=\linewidth]{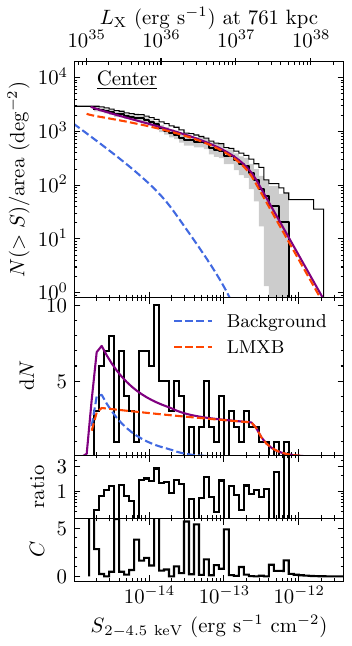}
    \end{minipage}
    \begin{minipage}{0.32 \textwidth}
    \centering
    \includegraphics[width=\linewidth]{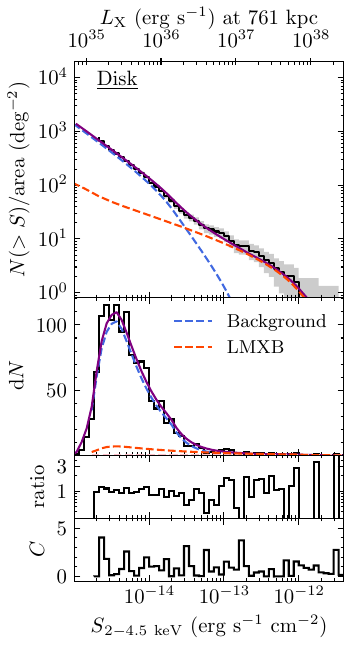}
    \end{minipage}
    \caption{The XLF fitting results for the background region (left), the center region (middle), and the disk region (right) between 2.0 {and} 4.5 keV. The background region combines sources from the north and south regions. 
    In the top panel of each region panel, the black line represents the incompleteness-corrected cumulative XLF, while the grey shaded area indicates uncertainties based on Gehrels' variance function \citep{1986ApJ...303..336G}. The green cross corresponds to the AGN XLF in {CDF-S} \citep{luoCHANDRADEEPFIELDSOUTH2016}. The blue and red dashed lines represent the fitted background and LMXB components, respectively. The purple solid line combines the background and LMXB components. 
    In the middle panel of the figures, the uncorrected distribution of the sources and the corresponding folded model are presented using the same color scheme as the top panel. The third panel is the ratio between model and observed data in middle panel. The bottom panel displays the residuals in C-statistic($C=2 \sum m_\mathrm{i}-\mathrm{N_i}+\mathrm{N_i} \mathrm{ln}(\mathrm{N_i}/m_\mathrm{i})$ with m and N as the model and the data,\citealt{Kaastra2017}). 
    Additionally, in the top panel for center region, the thin black stairs represent the XLF between $0-15^{\prime}$ for comparison.
    }
    \label{pic:Back_XLF}    
\end{figure*}

\subsection{Background regions: south and north}
\label{subsec:background}

In order to study the M31 source populations, we first need to account for the contributions from the background or foreground. We take the south and north region as the background region because of the minimal contributions from M31 sources.
We model the XLFs of the north and south region between 2.0-4.5 keV with the broken power law: 
\begin{equation}
 \mathrm{d}N/\mathrm{d}S= \begin{cases}
K \left(S / S_{\mathrm{ref}}\right)^{-\beta_{1}} & \left(S \leqslant f_{\mathrm{b}}\right) \\
K\left(f_{\mathrm{b}} / S_{\mathrm{ref}}\right)^{\beta_{2}-\beta_{1}}\left(S / S_{\mathrm{ref}}\right)^{-\beta_{2}} & \left(S>f_{\mathrm{b}}\right)
\end{cases}
\label{equ:brokenpl}
\end{equation}
where $S_{\mathrm{ref}}=10^{-14}\rm~erg~s^{-1}~cm^{-2}$, $f_\mathrm{b}$ is the break flux of the broken power-law model.
The fitting shows a slope of $\beta_2=2.7\pm 0.3$ and $3.0\pm0.3$ above the break flux in the north and south respectively. 
There is very little absorption due to neutral hydrogen between 2.0 and 4.5 keV. 
The slope in the North and South are consistent to the value of $2.72\pm0.24$ of dN/dS of AGN obtained by \cite{luoCHANDRADEEPFIELDSOUTH2016} in the Chandra Deep Field South, where the South is slightly steeper.

There is a subtle roll-over trend at the faint end of the XLFs in both the north and south regions, this trend becomes more pronounced within the XLF of the disk region (see Fig.~\ref{pic:LFv4}d and Fig.~\ref{pic:LF_model}~(left)) where deep observation is observed. The flux of this break is consistent with that of AGN XLF at approximately $5\times10^{-15}~\mathrm{erg~s^{-1}~cm^{-2}}$ \citep{luoCHANDRADEEPFIELDSOUTH2016} as indicated by the grey dashed line in the left panel of Fig.~\ref{pic:LF_model}.
The X-ray flux here is converted from the 2-7 keV to the 2-4.5 keV energy range, assuming the $\Gamma=1.7$ and $N_{\mathrm{H}}=6(\pm1)\times10^{20}~\rm cm^{-2}$, which is estimated from HI map by \citet{HI4PI2016}. 
This conversion factor is robust against variations in $N_{\mathrm{H}}$. 
For instance, for a source located behind the M31 disk where $N_{\mathrm{H}}$ 
can reach $\approx 3 \times 10^{21}$ cm$^{-2}$, the factor changes by a 
negligible $\lesssim 2\%$.

When comparing our results to the fiducial AGN model from the Chandra Deep Field-South (CDF-S) presented by \cite{luoCHANDRADEEPFIELDSOUTH2016} (see the left panel of Fig.~\ref{pic:LF_model}), we first examine the flux range around $10^{-14}~\mathrm{erg~s^{-1}~cm^{-2}}$. At this flux, the source densities in our North ($97\pm7~\mathrm{deg}^{-2}$ from 214 sources in $2.2~\mathrm{deg}^2$) and South ($84\pm7~\mathrm{deg}^{-2}$ from 149 sources in $1.8~\mathrm{deg}^2$) fields differ by $13~\mathrm{deg}^{-2}$. This difference is slightly larger than their combined Poisson error of $\sim10~\mathrm{deg}^{-2}$. This level of variation is, however, modest, as illustrated by a comparison with the CDF-S model: derived from a much smaller area ($0.13~\mathrm{deg}^2$), it predicts a density of $79\pm24~\mathrm{deg}^{-2}$.

However, a more significant discrepancy emerges at the faint end of the XLF. At fluxes approaching $10^{-15}~\mathrm{erg~s^{-1}~cm^{-2}}$, the AGN number density from the CDF-S model is $\sim30\%$ higher than what we expected in our background fields. This excess in the CDF-S at faint fluxes has also been noted by \cite{vulicXRaysBewareDeepest2016} in their study of the M31 disk. This finding strongly underscores the critical importance of establishing a suitable local background for modeling the XLF in the M31 disk, as relying on external deep-field models could introduce systematic errors. Furthermore, we note a slight excess in the North field relative to the South field, which is likely attributable to its lower Galactic latitude; this is discussed further in \S\ref{subsec:fgstar}.

To facilitate the subsequent analysis with higher statistical significance, we combine the north and south regions to form our background region. Without further decomposing into AGNs and Galactic sources, we employ a `AGN-like' broken power-law model (Equation.~\ref{equ:brokenpl}) to model our background XLF. However, as we exclude the identified background or foreground sources that brighter than $2\times10^{-13}\rm~erg~s^{-1}~cm^{-2}$, our background model is cut off beyond this flux.
The best-fit parameters for this model are as follows: $K_{\mathrm{back}}=(131 \pm 1) \times 10^{-14}\rm~deg^{-2}(erg~s^{-1}~cm^{-2})^{-1}$, $\beta_1=2.1\pm0.1$, $\beta_2=2.8\pm0.2$, and $f_\mathrm{b}=(1.8\pm0.6)\times10^{-14}\rm~erg~s^{-1}~~cm^{-2}$.
The result is also presented in the left panel of Fig.\ref{pic:Back_XLF} and summarized in Table.\ref{tab:XLF_fitting_parameter_summary}.

\subsection{Center region}

Next, we study the XLF in the center region ($R \leq 8'$, where $R$ is the distance to M31's center). In this region, there is a notable increase in the number density of sources compared to other nearby regions, and a distinct roll-over feature as shown in Figure~\ref{pic:LFv4}. The excess is obvious over all energy band and is consistent with previous findings (e.g., \citealt{kongChandraStudiesRay2003,Trudolyubov2002,vossStudyPopulationLMXBs2007}).
The XLF of M31's center region has been a subject of great interest in past decades. 
The XLF of the bulge region ($\lesssim 5'$) was first found to be flatter than the disk by \cite{kongXRayPointSources2002} and later confirmed by \cite{Williams2004} and \cite{vulicXRaysBewareDeepest2016}. This suggests a lack of bright HMXBs in the disk due to a low star formation rate and an aging population of LMXBs in the bulge, as discussed by \cite{vulicXRaysBewareDeepest2016}. Interestingly, this differs from studies of other spiral galaxies \citep{Colbert2004,Binder2012}. 
A more comprehensive summary of past surveys can be seen in \citealt{vulicXRaysBewareDeepest2016}.

It is believed that the sources in this region, which include the bulge of the spiral galaxy, are primarily contributed by LMXBs (e.g., \citealt{2002A&A...391..923G}).
The observed slope break occurring at approximately $2\times10^{-13}~\mathrm{erg\,~s^{-1}~cm^{-2}}$ or $1.4\times10^{37}\rm~erg~s^{-1}$ indicates a distribution transition of compact source of LMXBs
between neutron stars and black holes, as neutron star binaries have mass limit which also limit its Eddington luminosity and more luminous systems are black holes whose occurrence rates in the population are smaller \citep{fabbianoPopulationsXRaySources2006, Gilfanov2022}.

A rapid increase in the number of sources within $3^{\prime}$ of M31 \citep{vulicXRaysBewareDeepest2016}, the relatively lower spatial resolution of \emph{XMM-Newton} compared to \emph{Chandra}, and a distinct upward deviation at the bright end ($\sim 10^{-12}\rm~erg~s^{-1}~cm^{-2}$, see Fig.~\ref{pic:LF_model}, right) of the XLF in the center region suggest the presence of an overcrowding issue in our source detection.
In this scenario, multiple sources may be detected as a single source, leading to our detection of only 14 sources within $1^{\prime}$, while \emph{Chandra} detected 67 X-ray sources \citep{vulicXRaysBewareDeepest2016}. Additionally, the observed X-ray fluxes shift to higher luminosity, creating a distinct step feature in the XLF. To address this problem, we exclude sources within an inner $3'$ radius from our subsequent analysis as the number of the detected sources of \emph{Chandra} and \emph{XMM-Newton} are comparable \citep{vulicXRaysBewareDeepest2016}. As a result, most of the bulge region \citep{Courteau2011} has been excluded.
The new XLF between $3'{-}8'$ is displayed in the right panel of Fig.~\ref{pic:LF_model} and the middle panel of Fig.~\ref{pic:Back_XLF}. The XLF between $0'-8'$ is also shown in the thin solid line of Fig.~\ref{pic:Back_XLF} for comparison purposes. We model the XLF with a prefixed background component derived in \S\ref{subsec:background} and an additional LMXB component (Equation~\ref{equ:brokenpl}) folded with stellar mass function. The new XLF declines much faster, leaving the slope above the break as { $\beta_2=3.0\pm 1.6$}. Comparing with other LMXB results from external galaxies (e.g., $2.57^{+0.54}_{-0.28}$ from \citealt{Lehmer2019} and $2.06^{+0.06}_{-0.05}$ from \citealt{Zhang2012}), the slope is significantly steeper, indicating a deficit in bright sources. 
Our result suggests an aging population in the center $3^{\prime}{-}8^{\prime}$ region, more discussion can be seen in \S\ref{subsec:subregion}.

The other best-fit parameters are $K=(3.9\pm 0.5)\times10^{-12} \rm~deg^{-2}(erg~s^{-1}~cm^{-2})^{-1} $, $\beta_1=1.1\pm 0.1$, and the break flux $f_\mathrm{b}=(2.5 \pm 0.8) \times 10^{-13}\rm~erg~s^{-1}~cm^{-2}$ or the break luminosity is $L_{\mathrm{2.0-4.5 keV}}=(1.8\pm0.5)\times10^{37}\,\mathrm{erg\,s^{-1}}$. The value of the break luminosity is only a quarter of the reported value in \citealt{Lehmer2019}, but greater than that in \citealt{vulicXRaysBewareDeepest2016}, while it remains consistent with the findings of \citealt{Zhang2012}.

\subsection{Disk Region}
\label{subsec:disk}

\begin{figure*}
    \centering
    \includegraphics[width=0.99 \linewidth]{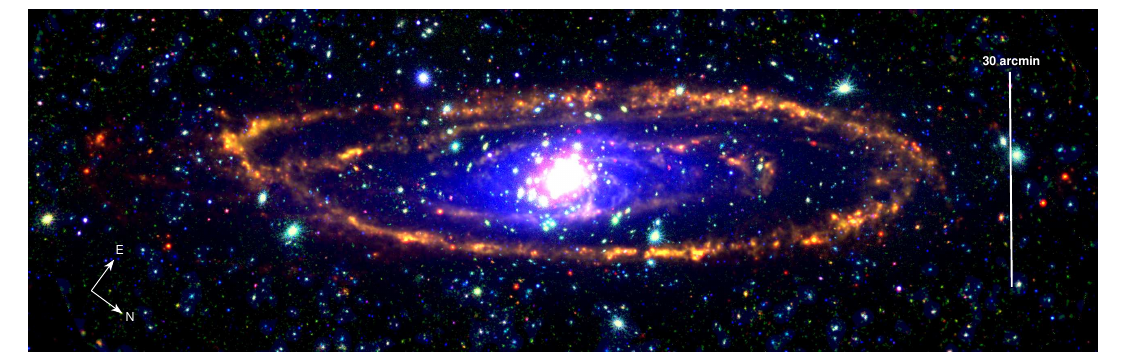}
    \begin{picture}(0,0)
    \put(-490,135){\makebox(0,0)[l]{{\color{white}X-ray \& IR}}}
    \put(-490,125){\makebox(0,0)[l]{{\color{red}Red: 0.5-1.0 keV \& 22 $\rm \mu m$}}}
    \put(-490,115){\makebox(0,0)[l]{{\color{green}Green: 1.0-2.0 keV \& 12 $\rm \mu m$}}}
    \put(-490,105){\makebox(0,0)[l]{{\color{cyan}Blue: 2.0-4.5 keV \& 3.4 $\rm \mu m$}}}
    \end{picture}
    \includegraphics[width=0.98 \linewidth]{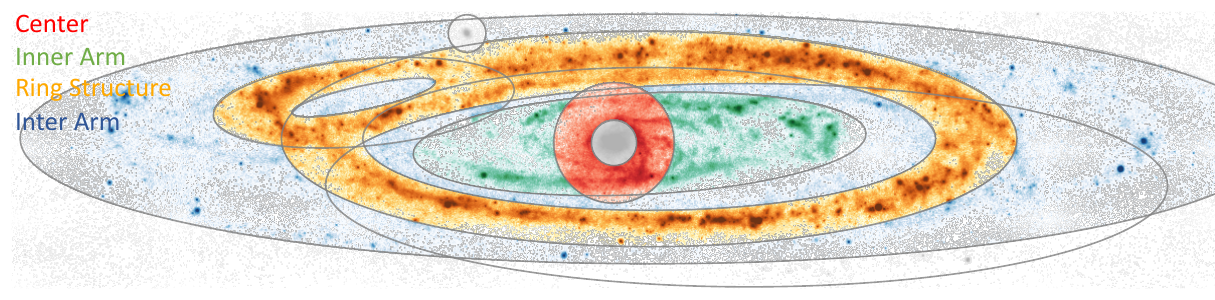}
    \caption{Top: the combined false color image of X-ray (XMM) and infrared (\emph{WISE}). The sources are X-ray detected, the centre blue diffuse feature comes from 3.4 $\rm \mu m$ image, and the yellow feature refer to the ring or arm structure traced by 22 and 12 $\rm \mu m$.  Bottom: \emph{WISE} W4 band image. `Center' (red): center 3-8 arcmin; `Inner Arm' (green): the inner arm region after excluding the center 8 arcmin; `Ring Structure' (orange): covering the ring feature. `Inter Arm' (blue): the combined inter-arm and outer-disk region.}
    \label{pic:disk_region_definition}
\end{figure*}

\begin{table*}
    \centering
    \begin{tabular}{c p{2.9cm} c cccccc}
        parameter  & unit & Background& Disk& {\small Center}& Inner Arm& Ring Structure&Inter Arm  &All\\
        \hline\hline
        area& $\rm~deg^2$& 4.70& 2.49& 0.05& 0.125&0.360&0.669 &1.19\\
 $M_{*}$ & $10^{10}\rm~M_{\mathrm{\odot}}$ & -& 6.5& 2.0& 1.8& 2.1&1.9&7.8\\
 SFR& $\mathrm{M_{\odot}~yr^{-1}}$& & 0.58& 0.05& 0.075& 0.32& 0.14&0.58\\
        \hline
        $K_{\mathrm{back}}$ &  $\rm 10^{-14}~(erg~s^{-1}~cm^{-2})^{-1}$ $\rm deg^{-2}$ & \textbf{$131\pm5$}&  131&  131&  131& 131&131&131 \\
        $\beta_1$ &  & $2.06\pm0.04$&  2.1&  2.1&  2.1& 2.1&2.1  &2.1  \\ 
        $\beta_2$ &  & \textbf{$2.9\pm0.1$}&  2.9&  2.9&  2.9& 2.9&2.9&2.9\\ 
        $f_\mathrm{b}$ & $10^{-14}\rm~erg~s^{-1}~cm^{-2}$& \textbf{$1.9\pm0.2$}&  1.9&  1.9&  1.9& 1.9&1.9&1.9\\  
        \hline
        $K_{\mathrm{LMXB}}$ & $\rm 10^{-14}~\mathrm{(erg~s^{-1}~cm^{-2})^{-1}} $ $(10^{10}~M_\odot)^{-1} $ &  -& $2.4\pm0.3$& \textbf{$9.3\pm1.3$}& \textbf{$5.9\pm4.5$}& \textbf{$12\pm8$}&\textbf{$2.7\pm1.4$}&$4.8\pm1.0 $\\
        $\beta_1$ &  &  -& \textbf{$1.17\pm0.03$}&      \textbf{$1.1\pm0.1$}&  \textbf{$1.2\pm0.5$}& \textbf{$0.8\pm0.4$}&\textbf{$1.3\pm0.2$}&$1.1\pm0.1$\\
        $\beta_2$ &  &  -& \textbf{$2.8\pm0.8$}&  \textbf{$3.5\pm1.1$}&  $3.9\pm3.4$& \textbf{$3.3\pm0.9$}&\textbf{$6$}  &$2.5\pm0.3$\\
        $f_\mathrm{b}$ &  $10^{-14}\rm~erg~s^{-1}~cm^{-2}$&  -& \textbf{$86\pm24$}&  \textbf{$25\pm7$}&  \textbf{$20\pm10$}& \textbf{$48\pm27$}&\textbf{$320\pm70$}&$28\pm7$\\
 C-statistic& & 27& 48& 64& 44& 57& 46&60\\
 expected 1$\sigma$ range& & $22\pm7$& $49\pm10$& $41\pm10$& $38\pm8$& $45\pm9$& $46\pm9$&$50\pm10$\\
 dof& & 91& 98& 98& 99& 99& 99&99\\
        \hline
        $\text{log}L_\mathrm{X,LMXB}$& $\rm~erg~s^{-1}$&  -&  $39.6\pm0.2$&  $39.15\pm0.11$&  $38.6\pm0.2$& $39.0\pm0.1$& $39.2\pm0.2$&$39.6\pm0.1$\\
        $\mathrm{log}~\alpha_{\mathrm{LMXB}}$&  $\rm ~erg~s^{-1}~M_{\odot}^{-1}$&  -&  $28.7\pm0.2$&  $28.85\pm0.11$&  $28.4\pm0.2$& $28.7\pm0.1$& $28.9\pm0.2$&$28.8\pm0.1$\\
        \hline
    \end{tabular}
    \caption{The best-fit parameters of the XLF modeling at the background, disk, `Inner Arm', `Ring Structure', `Inter Arm', and `All' region. The parameters with uncertainties are thawed parameters during fitting, while the others without uncertainties are fixed. $L_\mathrm{X,LMXB}$ is the luminosity of LMXB component within $\rm 0.5-8.0\,keV$, integrated between $\rm 10^{38}\,-\,10^{41}~erg~s^{-1}$. The  $\mathrm{log}~\alpha_{\mathrm{LMXB}}$ is in $\rm 0.5-8.0 \,keV$ which is converted from $\rm 2.0-4.5\,keV$ assuming absorped power law model with $\Gamma=1.7$ and $N_{\mathrm{H}}=6\times10^{20}~\rm cm^{-2}$. The expected $1 \sigma$ range of the C-statistic is derived from simulations. We generated mock datasets based on the best-fit model and subsequently re-fitted them with the identical model. The distribution of the C-statistic values from these fits determines the expected $1 \sigma$ range.\\}    \label{tab:XLF_fitting_parameter_summary}
\end{table*}

The XLF for the complete disk region is not thoroughly investigated in prior studies. Previous research (e.g., \citealt{kongXRayPointSources2002,vossStudyPopulationLMXBs2007,vulicXRaysBewareDeepest2016}) only cover part of the disk and claim potential distance-
dependence of the XLFs. Notably, \cite{stieleDeepXMMNewtonSurvey2011a} encompass the entire disk for the first time. However, they do not explore the XLF.

In the disk region, the M31 stellar sources are mixed with the background AGNs. We take the XLF fitting result  from the `background' region as our fixed background XLF component.
However, the study in the disk may suffer from the absorption issue.
Compared to the `background' region, the source number density in the disk increases with the energy range. In the 0.2-0.5 keV range, the number density is even lower than that in the `background' due to $N_{\mathrm{H}}$ absorption. Luckily, the absorption above 2.0 keV can be ignored.
Because the average $N_\mathrm{H}$ in the M31 disk region is approximately $3\times10^{21}\rm~cm^2$, compared to the foreground absorption $6\times10^{20}\rm~cm^2$,
it causes a shift in the X-ray source flux for a population with a photon index of 1.5-2.0 towards fainter values. Specifically, the shift amounts to around 1.5 dex, 0.4 dex, 0.1 dex, 0.03 dex, and 0.003 dex, for energy bands of 0.2-0.5 keV, 0.5-1.0 keV, 1.0-2.0 keV, 2.0-4.5 keV, and 4.5-12 keV, respectively. Even if the absorption is assumed to be 10 times higher ($3\times10^{22}\rm~cm^2$) than estimated, these shifts become 10~dex, 4.5~dex, 1~dex, 0.2~dex, and 0.025 dex for the corresponding energy bands.
The absorption below 1 keV is significant, but the 0.2~dex flux shift for 2.0-4.5 keV is marginal and has a minor impact on the observed XLF. Thus, in the following analysis we focus on the 2.0-4.5 keV result and use the background without the absorption correction.
However, the normalization of the XLF could be lower, for example $-2.7\times0.03$ dex for the AGN population, assuming $\mathrm{d}N/\mathrm{d}S\propto S^{-2.7}$.

The excess in the XLF in the disk compared to the `background' between 2.0-4.5 keV is attributed to M31 sources. By subtracting the `background', we obtain the XLF of M31 sources in the disk (see Fig.~\ref{pic:LF_model}, right). 
We estimate the number of M31 sources by integrating the background-subtracted XLF.
In our disk region, there are only about $80\pm30$ sources brighter than $5\times10^{35}\,\mathrm{erg\,s^{-1}}$, which is significantly smaller compared to our entire X-ray source catalog. Combined with $80\pm10$ sources from the center $3^{\prime}-8^{\prime}$ or $110\pm10$ from the center $0^{\prime}-8^{\prime}$, there are in total approximate 200 M31 sources brighter than $5\times10^{35}\,\mathrm{erg\,s^{-1}}$ included in our analysis.

\subsubsection{Subdivision of the Disk}
The overall XLF shape of the disk region (Fig.~\ref{pic:LF_model}, right) resembles that of the LMXB population in the Milky Way \citep{2002A&A...391..923G}, suggesting the disk region is also dominated by the LMXB population.
Despite this, M31 still has on-going, although only $0.6\rm~M_{\odot}~yr^{-1}$ \citep{Jarrett2019}, star formation activities to form HMXB population \citep{Williams2017PHAT,Lazzarini2021}. 
With the HMXB scaling relation from \citealt{Lehmer2019}, we estimate that the number of HMXBs to be detected with luminosity $L_\mathrm{0.5-8.0~keV}>$  $10^{36}~\mathrm{erg~s^{-1}}$ (or $f_\mathrm{2.0-4.5~keV}>4.6\times10^{-15}~\mathrm{erg~s^{-1}~cm^{-2}}$), denoted as $\mathcal{N}_\mathrm{HMXB}$, to be approximately {$65 \times \left(\frac{\text{SFR}}{M_{\odot}\,\text{yr}^{-1}}\right)$.
And considering the star-forming region usually follows the arm or ring structure, 
we want to further extract the XLF in the ring region of M31 where may have higher SFR comparing with other regions.

In particular, we divide the disk region into `Inner Arm', `Ring Structure', and `Inter Arm'. In Fig.~\ref{pic:disk_region_definition}, `Center' corresponds to the previous center 3-8 arcmin region where enclose most of the bulge, `Inner Arm' encompasses the inner spiral arm structure after excluding the central 8 arcmin, `Ring Structure' includes the most prominent dust ring structure of M31 with potential star formation activity, and `Inter Arm' combines the inter-arm region and the outer-disk region. The inter-arm region and the outer-disk region are combined for the limited number of sources in the two regions.
This subdivision approach not only helps decomposing the X-ray source population in M31, but also facilitates the comparison of populations across various regions of M31. Such sub-region analysis in M31 is not easily feasible for other external galaxies.

\begin{figure*}
    \centering
    \includegraphics[width=0.95 \columnwidth]{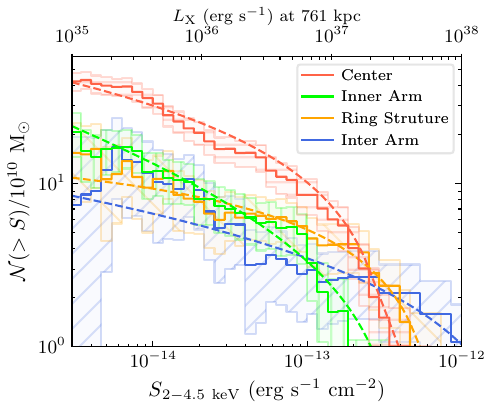}
    \includegraphics[width=0.95 \columnwidth]{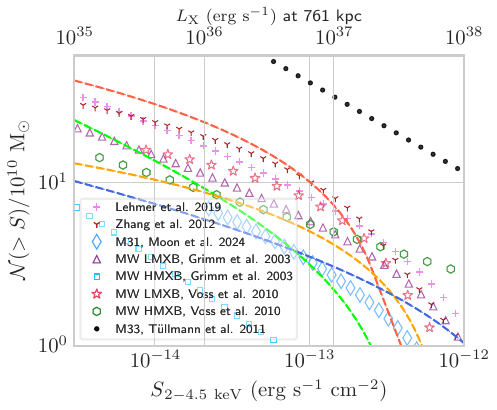}
    \caption{Left: The cumulative XLF in distinct regions (defined in Fig.~\ref{pic:disk_region_definition}). 
    Solid lines represent the cumulative XLF after subtracting contributions from the `Background' based on the fitting results. 
    Incompleteness corrections were applied using the stellar mass function. Dashed lines indicate the best-fit LMXB models for each region, shaded area represent 68\% uncertainty range. 
    Right: For comparison, our best-fit models (dashed lines) are shown alongside a compilation of XLF measurements from other studies. These other studies include: \citet{Lehmer2019} (violet plus symbols); \citet{Zhang2012} (brown tri-down symbols); M31 from \citet{Moon2024} (blue diamond symbols); Milky Way LMXBs from \citet{Grimm2003} (purple triangles); Milky Way HMXBs from \citet{Grimm2003} (light blue squares); Milky Way LMXBs from \citet{Voss2010} (red stars and green hexagons); and the M33 XLF from \citet{ChASeM332011} (filled black circles).
}
    \label{pic:background_subtracted_XLF}
\end{figure*}

Fig.~\ref{pic:background_subtracted_XLF} presents the background-subtracted XLFs for the `Center' through `Inter Arm' regions. While these XLFs show a close match to LMXB models \citep{Lehmer2019,Zhang2012,Grimm2003}, they do exhibit slight differences in their slopes and break fluxes. Crucially, they are distinct from the models dependent on the specific star formation rate (sSFR; the star formation rate per unit stellar mass) for the Milky Way \citep{Grimm2003,Voss2010} and M33 \citep{ChASeM332011}. 
The XLF of the `Inner Arm' has similar slope and break flux as in the `Center' (center 3-8$^{\prime}$). There is a slight trend that XLFs are progressively flatter and exhibit higher break flux when moving from `Inner Arm' to `Inter Arm'.  
Interestingly, the XLF shape in `Ring Structure' which encloses the ring structure, exhibits similar `broken' feature to that in `Center', which  covers the center 3-8 arcmin. This similarity suggests that the `Ring Structure' is still dominated by the LMXB population as in the center.
Conventionally, we associate HMXB with young populations in star formation regions, assuming they would be distributed throughout the arm \citep{fabbianoPopulationsXRaySources2006}. However, the recent star formation rate of M31 may be too low for us to decompose the HMXB population even in the region enclosing its ring structure. 
A detailed discussion of these source populations in the disk will be presented in \S\ref{subsec:M31SRC}.

\section{Discussion}\label{sec:Discussion}

\subsection{Galactic Sources}\label{subsec:fgstar}

In addition to the predominance of the AGN in both the north and south regions, Galactic sources (foreground sources in the Milky Way) also play a role in these regions. The number density of AGN is typically considered uniform given a sufficiently large sky coverage, whereas the number density of Galactic sources may vary depending on the distance to the Galactic plane.

The bright end of the XLF in the south ($\gtrsim10^{-13}\rm~erg~s^{-1}~cm^{-2}$) has more sources compared to the north, showing a significant difference particularly noticeable in the 0.2-0.5 keV range (Fig.~\ref{pic:LFv4}). This excess is due to six bright sources in the south region, in contrast to the north region which contains only a single such source.
Among these 6 sources in the south region, there are two very bright AGNs, two sources close to each other classified as supersoft sources (SSSs) in \citetalias{huangrui2023APJS}, and two unidentified bright sources. The single bright source in the north region is an AGN located at a significant projected distance ($\approx30~\mathrm{kpc}$) from the center of M31. Both of the SSSs have projected distances from the center of M31 exceeding 15 kpc, and one of them has been 
confirmed as a binary with a B type donor star that is 200 pc away from us.
Moreover, there are no known satellite galaxies of M31 within our field, except for M32 and M110. The M32 and M110 are not associated with these bright sources.
Therefore, the excess in the bright end of the cumulative XLF of the south region over the north region might be a result of statistical fluctuations and is unrelated to the surrounding environment of M31. The exclusion of these bright sources, as mentioned in Section \ref{subsec:note_on_bright_sources}, is thus crucial for analysis.

In addition to the bright sources, the XLFs of both the north and south regions are slightly higher than our fiducial AGN XLF model adopted from \citealt{luoCHANDRADEEPFIELDSOUTH2016} above 2 keV (see Fig.~\ref{pic:LF_model} and Fig.~\ref{pic:XLF_North_South}). 
This additional component likely arises from  Galactic sources, 
because of the low Galactic latitude of M31 ($b\approx-21.6^\circ$). 
Moreover, the faint end in the north region slightly surpasses that in the south region over all bands. This difference might be attributed to an increased presence of foreground stars in the north due to its proximity to the Milky Way plane.
The north and south regions are defined as being 30 arcmin away from the midplane, which is equivalent to 25 kpc away from the major axis after deprojecting the orientation, assuming a 0.73 ellipticity of the disk \citep{Courteau2011}. The probability of finding X-ray sources of M31 in these regions is relatively low. There are also no satellite galaxies within these fields, except for M32 and M110 which have no associated detected X-ray sources. 
However, we can not rule out the existence of the halo population of M31 in these regions.
Notably, there is at least one confirmed globular cluster (Bol 24, \citealt{Sargent1977}) located in the north region, approximately 26 kpc away from the center after de-projection.
The halo population in galaxy outskirt can reach even hundred kpc away from the galaxy as a result of kicking out (e.g., \citealt{Zhang2013}). However, considering the presence of several hundred sources in the north and south regions, this minor contribution is insufficient to establish a specific model component in our analysis. The potential contribution of this M31 X-ray halo population will be discussed in a dedicated spatial profile work.

\begin{figure*}
    \centering
    \begin{minipage}{0.32 \textwidth}
        \centering
        \includegraphics[width=\linewidth]{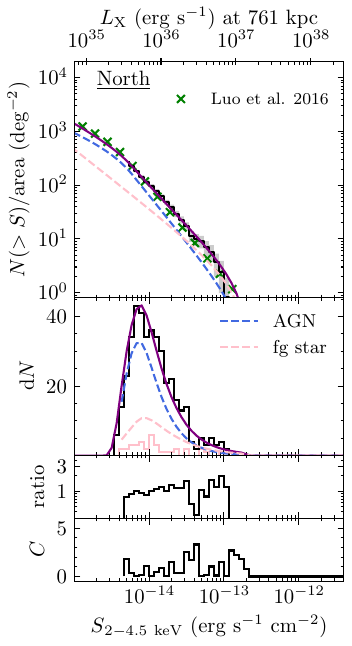}
    \end{minipage}%
    \begin{minipage}{0.32 \textwidth}
        \centering
        \includegraphics[width=\linewidth]{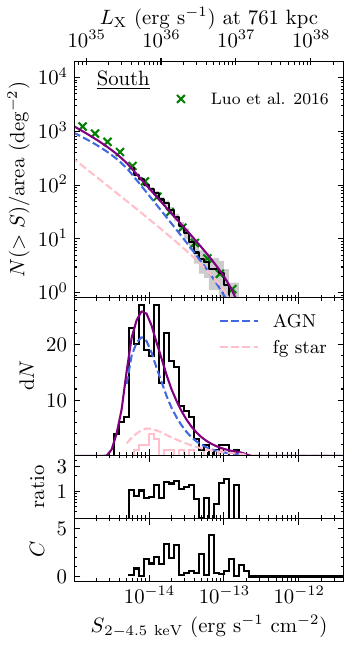}
    \end{minipage}
    \caption{Similar to Fig. 5, this figure focuses on the north region (left) and south region (right). The color scheme and symbols remain consistent, while the blue and pink dashed lines represent the AGN and Galactic component (mainly foreground stars), respectively. The purple solid line is the combination of the AGN and Galactic sources. An additional pink histogram is the classified Galactic sources in \citetalias{huangrui2023APJS}.}
    \label{pic:XLF_North_South}    
\end{figure*}

The X-ray luminosity functions (XLF) below 2 keV display striking consistency between northern and southern Galactic regions.
Spatial variations are constrained to be within $\lesssim 20\%$ at fluxes below $\mathrm{10^{-13}~erg~s^{-1}~cm^{-2}}$ in the 0.2–0.5 keV band, and within $\lesssim10\%$ in both the 0.5–1.0 keV and 1.0–2.0 keV bands.

(Fig.~\ref{pic:LFv4}a,b,c). However, average Milky Way foreground absorption ($N_{\mathrm{H}}$) in the north region ($\approx 6.7 \times 10^{20} \, \rm cm^{-2}$, \citealt{HI4PI2016}) is slightly higher than that in the south region ($\approx  5.0 \times 10^{20} \, \rm cm^{-2} $), due to its proximity to the Milky Way plane. Assuming an AGN photon index $\Gamma=1.7$ and an AGN XLF slope of 2.72 \citep{luoCHANDRADEEPFIELDSOUTH2016}, the higher absorption in the north leads to a 35\%, 10\%, and less than 1\% reduction in the normalization of the XLF in the 0.2-0.5 keV, 0.5-1.0 keV, and 1.0-2.0 keV energy bands when transitioning from the south region to the north region. Beyond 2 keV, this effect becomes negligible.
Intriguingly, the comparison of XLF in the 0.2-0.5 keV band (Fig.~\ref{pic:LFv4}), where significant absorption occurs, reveals that, apart from the previously discussed bright-end differences due to several identified sources, the northern region's XLF shows only a marginal ($\sim$20\%) deficit compared to the southern region at flux levels around $\mathrm{10^{-14}~erg~s^{-1}~cm^{-2}}$. More remarkably, the northern field actually contains 40\% more sources detected above $\mathrm{10^{-15}~erg~s^{-1}~cm^{-2}}$ than the southern field.
On one hand, we believe there are more sources in the north than presented, as they are shifted to lower X-ray flux, and these numerous additional sources may be attributed to foreground Galactic contributions. On the other hand, if they are foreground sources, {they would be less affected by absorption due to proximity as compared to sources in M31 or background AGN}. In higher energy bands, although there are still slightly more sources in the north than in the south, considering the decreasing impact of absorption, the contribution from foreground Milky Way sources is reducing. 
The similar count of dim sources in the lower energy range, along with a slightly higher count of sources in the north within the high energy range, implies that the majority of foreground Galactic sources are soft, like foreground stars. 

We then try to decompose the contribution of AGN and the Milky Way sources in the north and south regions (see Fig.~\ref{pic:XLF_North_South}) after {removing} the sources above $\rm 2\times10^{-13}~erg~s^{-1}~cm^{-2}$. We limit our analysis within 2.0-4.5 keV as foreground absorption can be ignored. 
There are 118 and 85 classified foreground stars in the north and south in \citetalias{huangrui2023APJS}, respectively.
However, only 36 in the north and 18 in the south were left after we remove the sources { that can still be detected in the single 2.0-4.5 keV band.}
This small amount is not only the result of a higher detection likelihood requirement in a single band, but also because the foreground stars are mainly soft. 
We fix the first slope of the AGN component above $10^{-14}\,\mathrm{erg\,s^{-1}\,cm^{-2}}$ to the result of \cite{luoCHANDRADEEPFIELDSOUTH2016} ($\beta_2=2.72$). 
The normalization is allowed to vary for the cosmic variance because the AGN XLF in CDF-S is higher around $5\times10^{-15}\,\mathrm{erg\,s^{-1}\,cm^{-2}}$ \citep{vulicXRaysBewareDeepest2016}.  
The Galactic component is represented by two power laws in two distinct regions, with their slopes interconnected.
The best-fit slope of the Galactic component is $2.2\pm 0.2$, while the normalization is $40\pm 30$ and $30\pm 30$ for the north and south regions, respectively. The normalization ratio between the north and south regions is slightly higher than the 20\% inferred from GAIA DR3 \citet{Gaia2023}.
The slope of XLF of Galactic sources is slightly higher than the slope of Galactic sources in CDF-S ($\alpha=1.88^{+0.36}_{-0.35}$, \citealt{luoCHANDRADEEPFIELDSOUTH2016}) and that of the Lockman hole by eROSITA ($\alpha=1.9$, \citealt{Belvedersky2022erosita}) which has a larger sample. 

{While Galactic sources contribute to the XLFs in both the north and south regions, with a stronger presence in the north, our north and south fields flank M31's disk. Their combined XLF (from Galactic sources and AGN) thus provides a good estimate of the total background relevant to M31. This overall background XLF resembles empirically a simple broken power law . We therefore treat this as a unified 'background' in our M31 analysis. This approach allows a clear focus on M31 sources, avoiding unnecessary and complex deconvolution of these other contributors.}

\subsection{M31 Sources}
\label{subsec:M31SRC}

\begin{figure*}
    \centering
    \includegraphics[width=0.33 \linewidth]{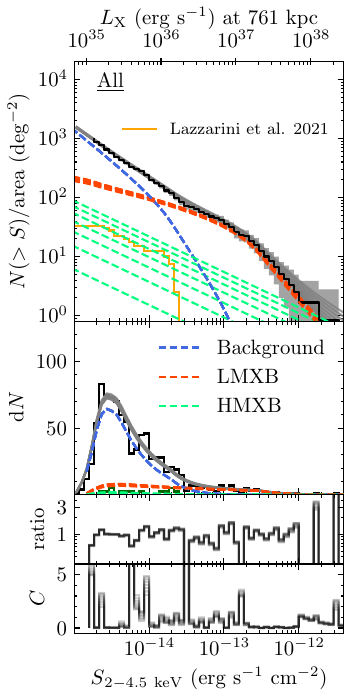}
    \begin{picture}(0,0)
    \put(-30,290){\makebox(0,0)[l]{{\color{black}(a)}}}
    \end{picture}
    \includegraphics[width=0.33 \linewidth]{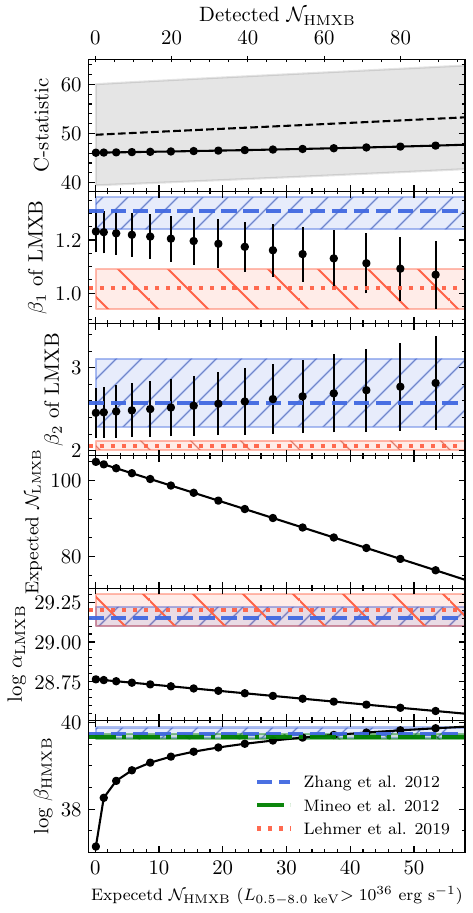}
        \begin{picture}(0,0)
    \put(-135,290){\makebox(0,0)[l]{{\color{black}(b)}}}
    \end{picture}
    \caption{(a): The XLF fitting in combined disk and center region. The color scheme and symbols remain consistent with Fig.~\ref{pic:Back_XLF}. The green dashed lines represent different HMXB contributions assuming the lowest 7 $N_\mathrm{HMXB}$ normalizations shown in (b) on the right. Additionally, we overlay the classified HMXBs (orange solid line) in \citealt{Lazzarini2021}, with its normalization elevated by a factor of 3 as the footprint of PHAT is roughly 1/3 of our `All' region. 
    (b): Fitting results for varying expected numbers of HMXBs with $L_\mathrm{0.5-8.0keV}>10^{36}\mathrm{erg~s^{-1}}$ in M31. The top x-axis shows the corresponding number of HMXBs that would be detected above our sensitivity limit, derived by convolving the intrinsic HMXB XLF model with the survey's incompleteness function of HMXB. From top to bottom, the panels display: the C-statistic of the fitting and the expected 1$\sigma$ range (grey shaded area) from simulation; the low-luminosity slope ($\beta_1$) of the LMXB population; the high-luminosity slope ($\beta_2$) of the LMXB population; the expected number of LMXBs with $L_\mathrm{2.0-4.5keV}>10^{36}\mathrm{erg\,s^{-1}}$; the ratio ($\alpha_\mathrm{LMXB}$, in $\mathrm{erg~s^{-1}M_{\odot}^{-1}}$) between the integrated LMXB luminosity ($L_{0.5-8\mathrm{keV}}>10^{36}\mathrm{erg\,s^{-1}}$) and stellar mass; and the ratio ($\beta_\mathrm{HMXB}$, in $\mathrm{erg\,s^{-1}(M_{\odot}yr^{-1})^{-1}}$) between the integrated HMXB luminosity ($L_{0.5-8\mathrm{keV}}>10^{36}\mathrm{erg~s^{-1}}$) and the star formation rate (SFR). The solid lines with points represent the fitting results for different fixed HMXB normalizations. For comparison, lines indicate the parameter values from previous studies: \citealt{Zhang2012} (red dotted), \citealt{Mineo2012HMXB} (green dash-dotted), and \citealt{Lehmer2019} (blue dashed). The correspondingly colored shaded regions represent the reported uncertainties associated with each of these parameter sets.}
    \label{pic:HMXB}    
\end{figure*}
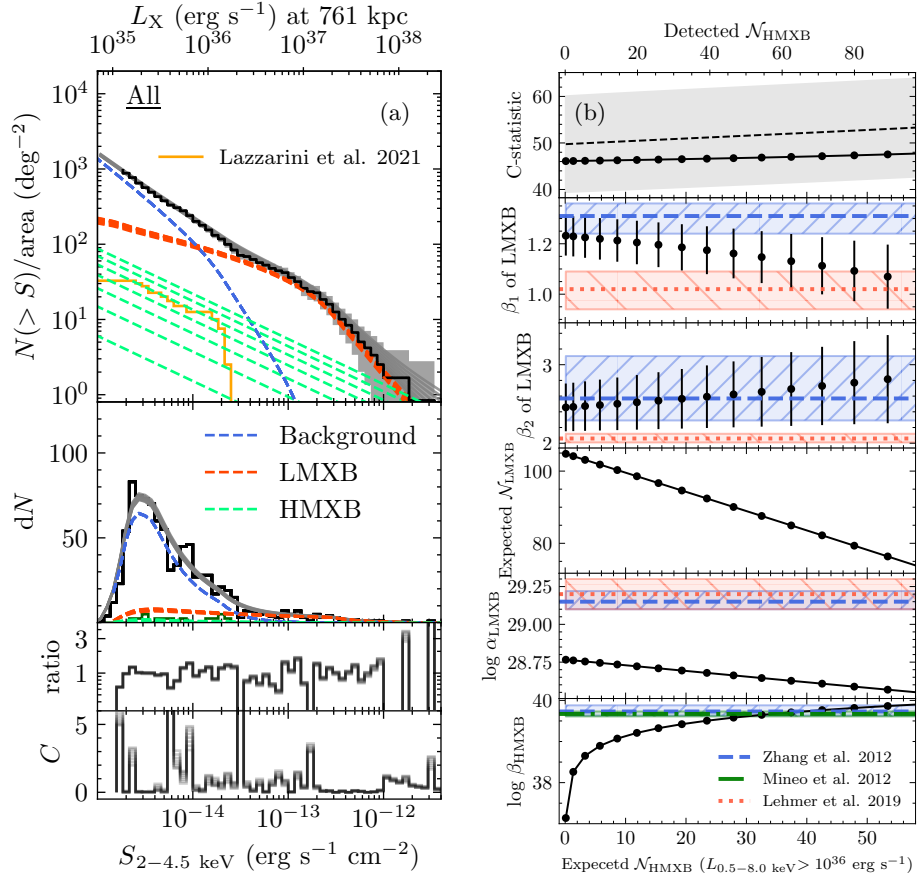

After establishing the `background' component of XLF, we can statistically study the distribution of M31 sources. 
To ensure that our analysis includes all potential M31 sources, we first combine the central and disk regions (denoted as `All'), excluding the central 3 arcmin due to source overcrowding.
Fig.~\ref{pic:background_subtracted_XLF} indicates that the XLF of remanining sources follows a broken power-law distribution, so we continue using broken power-law model (Eq.~\ref{equ:brokenpl}) which usually represents LMXB component, with a fixed background component constrained by the background region.
Given M31 has a low but non-negligible star formation rate ( $\approx 0.6~\mathrm{M_{\odot}~yr^{-1}}$), we expect to observe some HMXBs. Therefore, we also include an HMXB component, modeled as a power law (Eq.~\ref{equ:powerlaw}) with a fixed index $\gamma = 1.6$ (\citealt{Grimm2003}; \citealt{Mineo2012HMXB})
\begin{equation}
 \mathrm{d}N/\mathrm{d}S = K  (S / S_{\mathrm{ref}})^{-\gamma} \\
\label{equ:powerlaw}
\end{equation}
We fit the X-ray flux distribution of our sources, $\mathrm{d}N(S)$, shown in the middle panel of Fig.~\ref{pic:HMXB}a, using three components: the `background' component, the LMXB component, and the HMXB component. The `background' component is the combination of background and foreground sources (i.e., AGN, forground star) and fixed at the results in Background region. To account for incompleteness in the observed distribution, we convolve the background component with the sky coverage function, the LMXB component with the stellar mass function, and the HMXB component with the SFR function.

Preliminary fitting suggests that there are $114\pm13$ M31 sources among the 240 sources that are brighter than $\rm 10^{-14}~erg~s^{-1}~cm^{-2}$ in the `All' region's XLF. Meanwhile, the fitting prefers an absence of HMXBs when HMXB normalization is allowed to vary freely. 

To further constrain the contribution of HMXBs, we fix the HMXB normalization at several values and fit the remaining components (see Fig.~\ref{pic:HMXB}a). This allows us to evaluate the impact of the HMXB component on the overall fit. Based on the C-statistic of the fitting (top panel of Fig.~\ref{pic:HMXB}b), the fit tends to improve slightly with fewer HMXBs, as indicated by the lower C-statistic values. However, the change in the C-statistic is relatively small compared to its expected range (shaded region on the top panel of Fig.~\ref{pic:HMXB}b), suggesting that the measured XLF cannot strictly constrain the HMXB component.
Meanwhile, the slope of the LMXB component at high X-ray fluxes (\(\beta_2\)) remains largely unaffected by the inclusion of more HMXB. In contrast, the slope at lower X-ray fluxes (\(\beta_1\)) and the total number of LMXBs decrease as more HMXBs are assumed. Despite these adjustments, our fitting still weakly favors fewer HMXBs overall.

Comparing our results with previous studies by \citealt{Zhang2012}, \citealt{Mineo2012HMXB}, and \citealt{Lehmer2019}, whose parameter values are indicated by lines and their reported uncertainties by the correspondingly colored shaded regions (red, green, and blue, respectively) in Fig.~\ref{pic:HMXB}b, 
we find that our measured $\alpha_\mathrm{LMXB} = L_\mathrm{LMXB}/M_{*}$ is much lower. If we assume that the $\beta_\mathrm{HMXB}=L_\mathrm{HMXB}/SFR$ is consistent with theirs, we would expect $\approx 35$ HMXBs in M31 above $L_\mathrm{0.5-8.0~keV}=10^{36}~\mathrm{erg~s^{-1}}$. 
A recent effort by \citet{Lazzarini2021} to identify HMXBs in M31, using the Chandra-PHAT survey by \citet{Williams2018}, reported approximately 20 HMXB candidates within the PHAT footprint. To compare, we overlay the XLF of matched HMXBs from our catalog in Fig.\ref{pic:HMXB}a (scaled by a factor of three, as the Chandra-PHAT survey covers roughly one-third of our 'All' region). The observed HMXB luminosity function shows a clear break just above $10^{36}$ erg/s and notably lacks bright sources above $\rm 3\times10^{-14}~erg~s^{-1}cm^{-2}$, hinting a rapid decline in the brightness of the HMXB population. The cumulative XLF of the classified HMXBs aligns closely with our fifth HMXB model in Fig.~\ref{pic:HMXB}a, yielding $\beta \approx 10^{39.3}~\rm erg~s^{-1}(M_{\odot}~yr^{-1})^{-1}$, which is still lower than that reported in \citet{Mineo2012HMXB} and \citet{Lehmer2019}. 
While the observed break in the HMXB luminosity function could potentially be modeled with a broken power law rather than the single power law employed in our analysis, this would not significantly affect our conclusions regarding $\alpha_\mathrm{LMXB}$. This is because these classified faint HMXBs in \citealt{Lazzarini2021} exhibit only $\beta \sim 10^{20}\rm~erg~s^{-1}(M_{\odot}~yr^{-1})^{-1}$, total X-ray luminosity is dominated by the brightest sources, which would remain classified as LMXBs even with higher HMXB contributions. As demonstrated in Fig.~\ref{pic:HMXB}b, varying the HMXB contribution causes only modest changes to $\alpha_\mathrm{LMXB}$ (less than 0.2 dex), insufficient to reconcile our results with previous literature values.

On one hand, the specific star formation rate of M31, $\rm log(sSFR/yr^{-1}$) $\approx -11.1$, suggests an LMXB-dominated environment \citep{Gilfanov2022}.
On the other hand, by combining the X-ray binary luminosity evolution model from \citealt{Fragos2013} with the star formation history (SFH) of M31's disk from the PHAT survey (\citealt{Williams2017PHAT}), we estimate that HMXBs contribute less than $3\%$ of the total LMXB luminosity. Apply this fraction to our measured $L_\mathrm{LMXB}$, then $\beta_\mathrm{HMXB}$ is less than $10^{38.4}~\mathrm{erg~s^{-1}(M_{\odot}~yr^{-1})^{-1}}$. More detail of the estimation can be seen in \S\ref{subsec:SFH}.
Therefore, although we can not strictly constrain the HMXB contribution directly with our XLF, we conclude that it is minimal enough to be excluded from our analysis.
Even if we consider significant HMXB contributions, they would not have much impact on our discussion of LMXBs, as they do not significantly alter the XLF shape of LMXBs (see red dashed line in Fig.~\ref{pic:HMXB}b or the values of $\beta_1$ and $\beta_2$ in the right panel) and they only reduce $\mathrm{log}\alpha_\mathrm{LMXB}$  less than 0.2 dex, still leaving a huge gap (0.4 dex) between our results and the values reported in the literatures (see $\alpha_\mathrm{LMXB}$ in Fig.~\ref{pic:HMXB}b).

\subsubsection{M31 Sources in the Sub-regions}
\label{subsec:subregion}

\begin{figure*}
    \centering
    \begin{minipage}{0.32\textwidth}
        \centering
        \includegraphics[width=\linewidth]{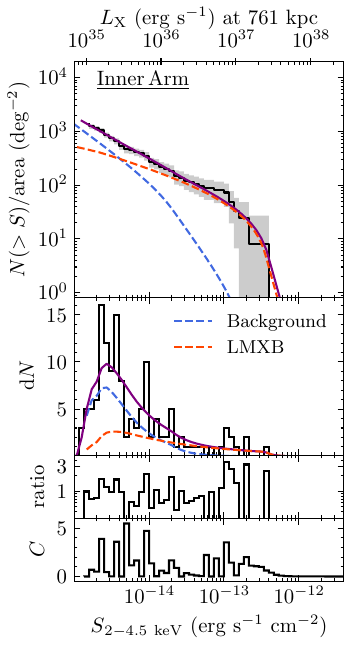}
    \end{minipage}%
    \begin{minipage}{0.32\textwidth}
        \centering
        \includegraphics[width=\linewidth]{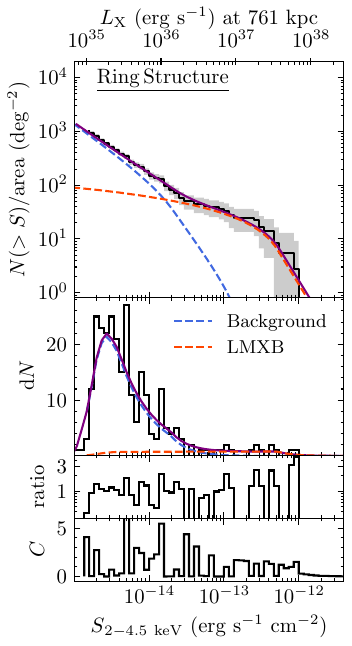}
    \end{minipage}
    \begin{minipage}{0.32\textwidth}
        \centering
        \includegraphics[width=\linewidth]{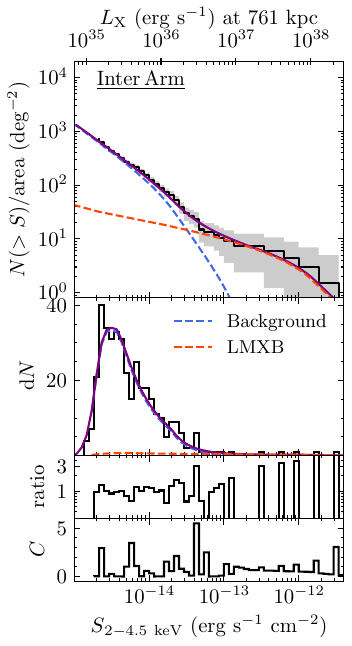}
    \end{minipage}
    \caption{The XLF fitting results for sources in the Inner Arm region (left), the Ring Structure region (middle), and the Inter Arm region (right) between 2.0 and 4.5 keV.  The color scheme and symbols are consistent with their definitions in Fig.~\ref{pic:Back_XLF}.}
    \label{pic:XLF_arm_disk}    
\end{figure*}

As described in \S\ref{subsec:disk}, we divide the disk into three different regions: `Inner Arm' (inner-arm), `Ring Structure' (ring), and `Inter Arm' (inter-arm and outer-disk). Additionally, `Center' represents the previously defined central 3–8 arcmin region. 
We use the modeling method from \S\ref{subsec:M31SRC} to fit these regions, without HMXB component. The fitting parameters are summarized in Table~\ref{tab:XLF_fitting_parameter_summary}. While the background model parameters are fixed, we allow the XLF model parameters for LMXBs to vary between regions. 
Generally, the results in sub-regions are similar to the features observed in the `All' region. 
While LMXBs dominate the source population, we still see significant differences in XLF shapes across the regions.

In `Center', we observe a notable lack of bright X-ray sources and a steep slope in the XLF. Specifically, our result reveals a lower break flux within the broken power-law model, a steeper slope compared to the result in \citealt{Zhang2012} (see magenta dashed line in the middle panel of Fig.~\ref{pic:Back_XLF}), and consequently a lower integrated luminosity per unit stellar mass. 
This steep slope, combined with the lower integrated luminosity per unit stellar mass, suggests an aging population of LMXBs in this inner region. As noted by \citet{Zhang2011}, LMXB populations tend to evolve with galaxy age, where younger galaxies typically display a higher abundance of bright sources and fewer faint ones per unit of stellar mass. For instance, in younger galaxies, the XLF of LMXBs often extends beyond a luminosity threshold of approximately $10^{39}~\text{erg}~\text{s}^{-1}$, a feature less commonly observed in older galaxies like M31.

In `Inter Arm', however, the sharp decline in the XLF above the break flux cause the high-energy slope parameter ($\beta_2$) to approach our upper limit. To address this, we freeze $\beta_2$ at a value of 6, as further increases to this value do not significantly alter the XLF shape. The break flux $f_\mathrm{b}$ in `Inter Arm' approaches the high-end X-ray flux limit of the XLF, making $\beta_2$ strongly dependent on this limit. As a result, comparisons between `Inter Arm' and the other regions require caution. Alternatively, a single power-law model with an index of 1.4 also provides a good fit for the Inter Arm region, yielding a C-statistic of 47 (expected $1\sigma$ range: $49 \pm 9$). Nonetheless, we observe that `Inter Arm' contains a higher number of bright sources.

When comparing these subregions, `Center' exhibits the highest number of M31 sources per stellar mass among four subregions. The XLF in `Inner Arm' follows a similar broken power-law pattern to `Center', while the XLFs in `Ring Structure' and `Inter Arm' appear notably flatter. Moving from `Inner Arm' to `Inter Arm' (Fig.~\ref{pic:background_subtracted_XLF}, left), we observe a clear trend of increasing break flux and a greater abundance of bright sources.
Given that \citet{Zhang2012} and \citet{Lehmer2014} find younger LMXB populations tend to produce XLFs with flatter slopes, our observed trend may suggest that the outer regions of M31 probably contain a younger population of LMXBs compared to the inner regions. 
This distribution pattern indicates a difference in evolution history between the inner and outer regions of M31.

Several studies of XRB emission from galaxies in the nearby universe ($D\lesssim$50 Mpc) have established that the LMXB and HMXB XLFs and the integrated emission are correlated with the stellar mass and SFR, respectively \citep{Grimm2003, gilfanovLowmassXrayBinaries2004, Mineo2012HMXB, Zhang2012, Lehmer2019}. These correlations have been widely considered as `universal' beyond the scope of XRB-specific studies. 
We therefore overlay the total luminosity per unit SFR ($L_\mathrm{X}/SFR$) of our different M31 regions in Fig.~\ref{pic:Luminosity_sSFR}, which is from Fig.~6 of \citealt{Lehmer2019}. The best-fit global model (black solid curve) represents the general correlation between integrated X-ray source emission and both stellar mass and SFR for galaxies by \citet{Lehmer2019}. 
However, our results for M31 show a systematic deviation from this global relationship. 
Among these regions, `Inner Arm' appears notably dimmer, while `Inter Arm' is slightly brighter. The discrepancy with the global $\alpha_\mathrm{LMXB}=L_\mathrm{LMXB}/\mathrm{M_{*}}$, along with the regional inconsistencies in $\alpha_\mathrm{LMXB}$ between our regions, suggests that factors beyond stellar mass and SFR are shaping the X-ray luminosity across different regions. This observed luminosity deficit is also rooted in the specific shape of M31's LMXB XLF, which we compare to that of the Milky Way and other nearby external galaxies in the right panel of Fig.~\ref{pic:background_subtracted_XLF}. While the cumulative XLF per stellar mass in M31 appears comparable to that of these galaxies at lower luminosities, $\alpha_\mathrm{LMXB}$ is dominated by its brightest members. Therefore, the pronounced deficit of the most luminous sources in M31, evidenced by its steeper XLF slope above a higher break luminosity ($>10^{37.5}~\mathrm{erg~s^{-1}}$), directly explains its lower integrated luminosity per unit stellar mass ($\alpha_{\rm LMXB}$). Beyond these simple correlations, X-ray binary population synthesis models (e.g., \citealt{Fragos2013}) indicate that the number of X-ray binaries, the shape of their luminosity function, and their integrated X-ray luminosity evolve strongly as a function of both stellar age and metallicity. Generally, as a stellar population ages, the integrated X-ray luminosity of its X-ray binaries declines, and lower metallicity populations are associated with higher integrated X-ray luminosities.
To gain preliminary insights into how these factors might affect M31 and understand the observed $L_\mathrm{X}/\mathrm{M_\mathrm{\odot}}$, we use the SFH of M31's disk, combined with simple X-ray binary evolution models, as a {{conceptual exploration}}. This approach provides a tentative basis for interpreting the observed $L_\mathrm{X}/\mathrm{M_\mathrm{\odot}}$ variations across different regions of M31.

\subsubsection{Impact of Star Formation History on X-ray Source Populations}
\label{subsec:SFH}

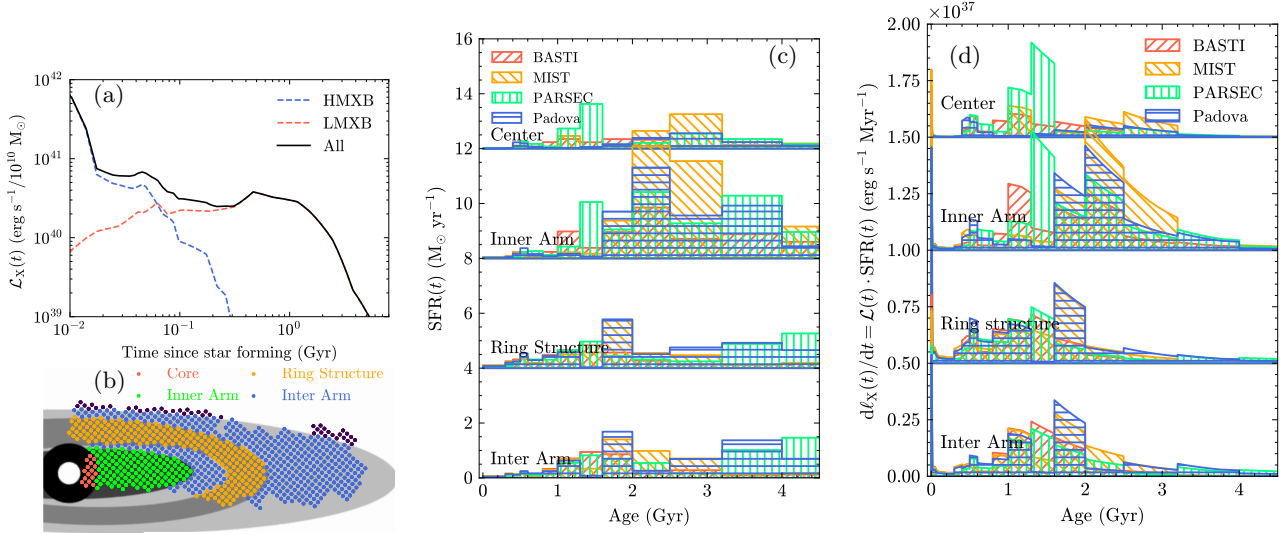
\begin{figure*}
    \centering
    \includegraphics[width=0.29 \linewidth]{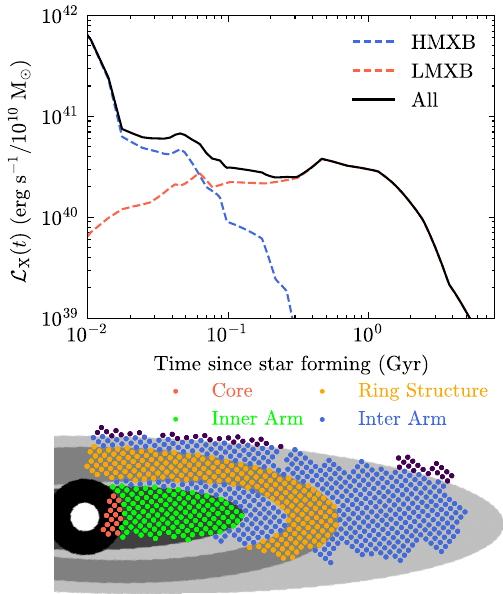}
    \begin{picture}(0,0)
    \put(-117,165){\makebox(0,0)[l]{{\color{black}(a)}}}
    \put(-117,57){\makebox(0,0)[l]{{\color{black}(b)}}}
    \end{picture}
    \includegraphics[width=0.3 \linewidth]{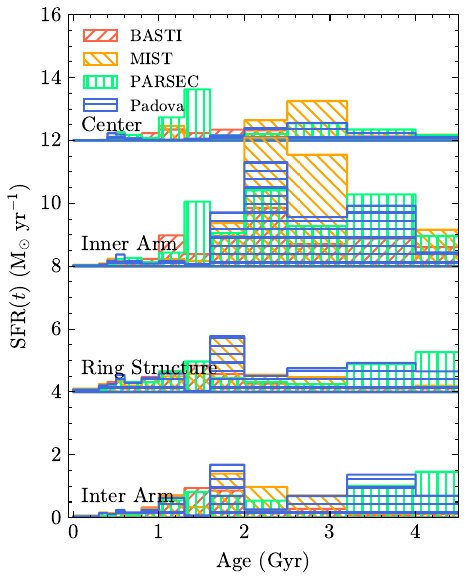}
    \begin{picture}(0,0)
    \put(-25,180){\makebox(0,0)[l]{{\color{black}(c)}}}
    \end{picture}
    \includegraphics[width=0.32 \linewidth]{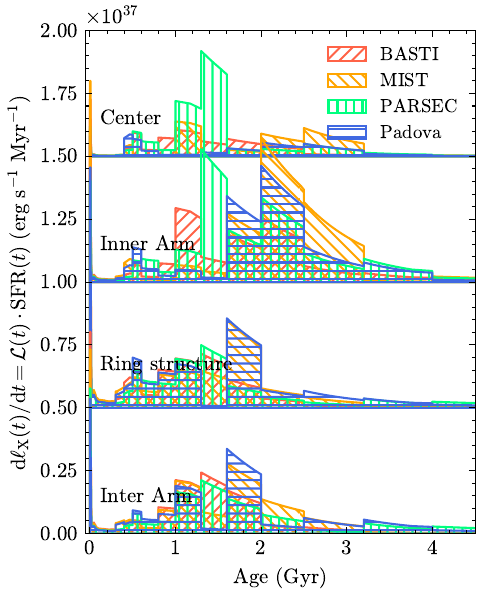}
    \begin{picture}(0,0)
    \put(-130,180){\makebox(0,0)[l]{{\color{black}(d)}}}
    \end{picture}
    \caption{(a): Luminosity evolution of an X-ray binary (XRB) population with solar metallicity, modeled under a single-burst star formation scenario. The blue dashed line represents the contribution from high-mass X-ray binaries (HMXBs) which are only bright within 100$-$300 Myrs, the red dashed line from low-mass X-ray binaries (LMXBs) which gradually fade after 1 Gyr, and the black solid line indicates the total luminosity from both populations. Figure adapted from \citealt{Fragos2013, Lehmer2024}; (b): Spatial distribution of subregions within the PHAT survey of M31, with color-coded dots representing `Center' (red), `Inner Arm' (green), `Ring Structure' (orange), and `Inter Arm' (blue). 
    (c): Star formation histories (SFH) of each region, vertically shifted for clarity. SFH models based on different stellar evolution tracks (BASTI, MIST, PARSEC, and Padova) are shown in red, orange, green, and blue, respectively. The recent SFH indicates a quiescent phase following a starburst that occurred approximately 1 Gyr ago. 
    (d): The differential X-ray luminosity contribution ($\mathrm{d}L_\mathrm{X}/\mathrm{d}t$) of each past star formation epoch to the \textit{total present-day} X-ray luminosity. Each point on the curve is calculated by multiplying the SFR at a given lookback time $t$ (from panel c) by the corresponding value from the binary luminosity evolution model $L_X(t)/M_*$ (from panel a), as described in Equation~\ref{equ:dL/dt}. The sloped tops within each bin result from multiplying the constant SFR value by the continuously decreasing evolution model over that time interval. The total present-day luminosity for a region is the integral of its entire curve in this panel. The color scheme and model tracks in d are identical to those in c. In `Inner Arm', There are more contributions from the stellar populations older than 2 Gyr.}
    \label{pic:SFH}
\end{figure*}

\begin{figure}
    \centering
    \includegraphics[width=0.95 \linewidth]{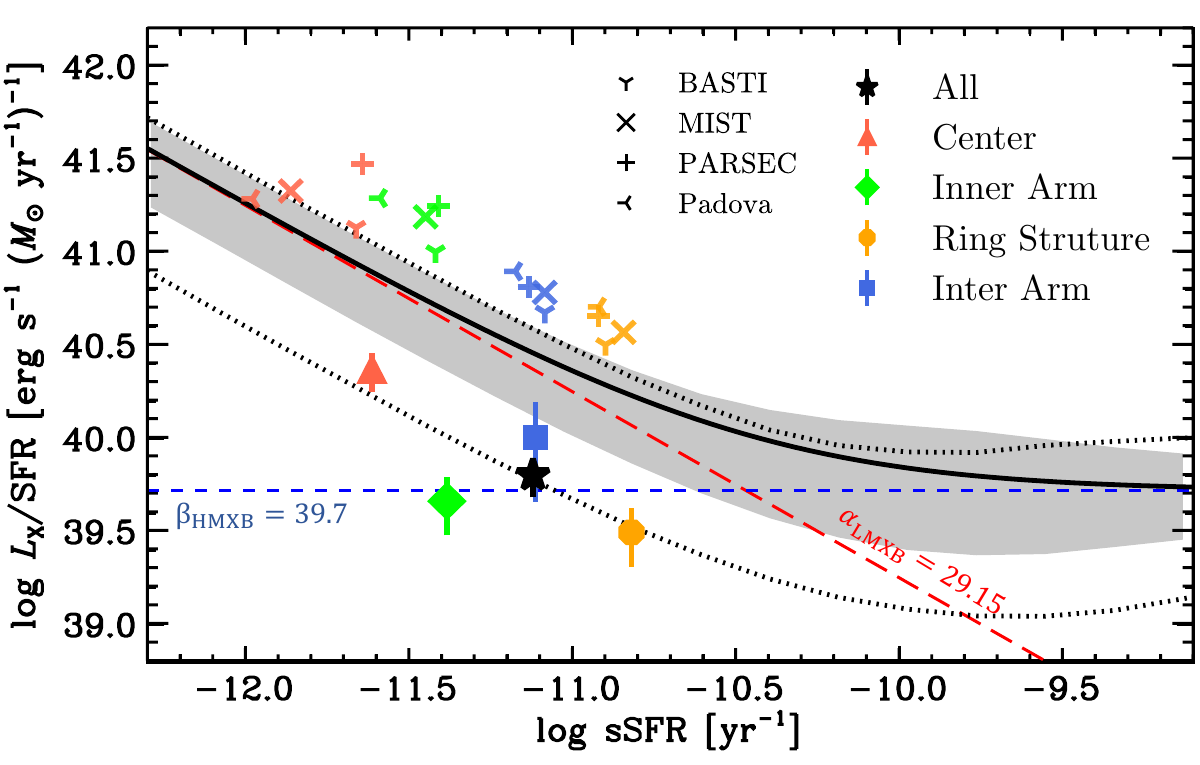}
    \caption{Integrated X-ray luminosity per unit star formation rate ($L_\mathrm{X}$/SFR) vs. specific star formation rate (sSFR) for regions of M31. 
    Measurements for regions within M31 are overlaid, with the `All' region (black star) representing the combined region of `Center' (red triangle), `Inner Arm' (green diamond), `Ring Structure' (orange circle), and `Inter Arm' (blue square). 
    Predicted $L_\mathrm{X}$/SFR values for each region are overlaid using the same colors as the measured data [BASTI (star), MIST (cross), PARSEC (+), and Padova (arrow)], derived from the SFH (Fig.~\ref{pic:SFH}c) from \citealt{Williams2017PHAT} and the X-ray binary luminosity evolution model (Fig.~\ref{pic:SFH}a) from \citealt{Fragos2013}. The stellar mass estimated from SFH is scaled to the survived mass.
    The luminosity is integrated between $10^{36}$ and $10^{41}~\rm erg~s^{-1}$. The energy range for luminosity has been converted from 2.0–4.5 keV to 0.5–8 keV for comparison purposes. 
    {Lines are adapted from Fig.6 of \citealt{Lehmer2019}; The solid black curve is the combined HMXB+LMXB prediction from \citealt{Lehmer2019}, fitted across 21 galaxy bins. 
    Gray shading and black dotted lines respectively indicate the 1$\sigma$ scatter at galaxy samples with median mass of $\rm 2\times10^{10}~M_{\odot}$ and $\rm 3\times10^{9}~M_{\odot}$,
    more details can be found in that paper.}
    The blue dashed line indicates the expected $L_\mathrm{X}$/SFR for high-mass X-ray binaries (HMXBs) with $\log \beta_\mathrm{HMXB} = 39.7~\mathrm{erg\,s^{-1}(M_{\odot}~yr^{-1})^{-1}}$, while the red dashed line represents the expected value for low-mass X-ray binaries (LMXBs) with $\log \beta_\mathrm{LMXB} = 29.15$. 
    The black solid line shows the combined contribution from both LMXBs and HMXBs. 
    Predicted $L_\mathrm{X}$/SFR values for each region are overlaid using the same colors as the measured data, based on the SFH (Fig.~\ref{pic:SFH}c) from \citealt{Williams2017PHAT} and the X-ray binary luminosity evolution model (Fig.~\ref{pic:SFH}a) from \citealt{Fragos2013}. The stellar mass estimated from SFH is scaled to the survived mass.}
    \label{pic:Luminosity_sSFR}
\end{figure}

Theoretical luminosity evolution models $\mathcal{L}(t)$ based on binary population synthesis, such as those proposed by \citet{Fragos2013}, suggest that LMXBs gradually fade over 1–2 Gyr after a star formation event, while HMXBs are short-lived, typically emitting strongly for only around 100 Myr (see Fig.~\ref{pic:SFH}a). 
To explore how these population evolve with M31's unique SFH, we use $\mathrm{SFR}(t)$ data from the PHAT survey (Fig.~\ref{pic:SFH}b and c, \citealt{Williams2017PHAT}) as a preliminary basis for analysis.  \citet{Williams2017PHAT} provide the $\mathrm{SFR}(t)$ for their 826 subregions, each measuring $83^{''}\times83^{''}$. We rebin these subregions into our four defined regions, as shown in Fig.~\ref{pic:SFH}b.
According to \citet{Williams2017PHAT}, M31's SFH has several distinct epochs: a period of intense star formation over 8 Gyr ago, a relatively quiescent phase until around 4 Gyr ago, another significant burst about 2 Gyr ago, and then a return to relative quiescence. 
Fig.\ref{pic:SFH}c shows the SFH in the overlapping regions (Fig.~\ref{pic:SFH}b) between the PHAT survey and our regions.
The `Inner Arm', in particular, experienced relatively intense star formation around 2–4 Gyr ago compared to the other regions. 
The relative contribution of different age populations to current X-ray luminosity ($\mathrm{d}\ell_X(t)/\mathrm{d}t$, Eq.~\ref{equ:dL/dt})} is shown in Fig.\ref{pic:SFH}d, where `Inner Arm' {has significant contributions from the 2–3 Gyr population. 
\begin{equation}
\mathrm{d}\ell_X(t)/\mathrm{d}t = \mathcal{L}(t)\cdot dM =
\mathcal{L}(t)\cdot\text{SFR}(t)
\label{equ:dL/dt}
\end{equation}
\begin{equation}
L_{X} = \int_0^{T_{\text{age}}} \frac{\mathrm{d}\ell_X(t)}{\mathrm{d}t}\mathrm{d}t = \int_0^{T_{\text{age}}} \mathcal{L}(t) \cdot \text{SFR}(t)  dt
\label{equ:L}
\end{equation}

Based on Eq.~\ref{equ:L} and Fig.~\ref{pic:SFH}d, we derive the $L_\mathrm{X}/\mathrm{SFR}$ in different regions and overlay them on Fig.~\ref{pic:Luminosity_sSFR}. This preliminary model shows several interesting features. First, the sSFRs derived from the SFH agree well with those estimated from \emph{WISE} imaging. Second, the predicted $L_\mathrm{X}/\mathrm{SFR}$ values are consistently higher than our measured values and the expectation of \citealt{Lehmer2019}. To facilitate a direct comparison with observational data, the luminosity evolution model ($\mathcal{L}(t)$) in Fig.~\ref{pic:SFH}a approximates the 0.5-8.0 keV luminosity as half of the bolometric value from \citet{Fragos2013}. This conversion is empirically supported by recent frameworks (e.g., \citealt{Lehmer2024}) and makes our comparison with the \citet{Lehmer2019} data more appropriate. Third, the trend of $L_\mathrm{X}/\mathrm{SFR}$ vs. sSFR in Fig.~\ref{pic:Luminosity_sSFR} indicates that LMXBs are likely the primary contributors to X-ray luminosity across all regions, including `Ring Structure', which contains the prominent dust ring of M31. The contribution of HMXBs in the preliminary model is less than 3\% of the total luminosity, supporting the conclusion in \S\ref{subsec:M31SRC} that M31's X-ray emission is dominated by LMXBs.

When we compare our measurement with the model predicted values in Fig.~\ref{pic:Luminosity_sSFR}, we observe that `Inner Arm' appears relatively faint, while `Inter Arm' is slightly brighter. As shown in Fig.~\ref{pic:SFH}c, `Inner Arm' has been quiescent over the past 1 Gyr but underwent significant star formation {between 2 and 4 Gyr ago}. This past activity likely accounts for much of `Inner Arm''s current X-ray luminosity. Notably, the enhanced star formation around 2–3 Gyr ago distinguishes `Inner Arm' from other regions in our study. Since the metallicity across all regions is approximately in solar metallicity, and `Inner Arm' actually has a larger fraction of low-metallicity stellar sources, which would typically make it brighter \citep{Williams2017PHAT}, the observed dimness of `Inner Arm' is unlikely to be due to metallicity effects. Furthermore, adjusting the X-ray luminosity model for different metallicities only shifts the overall brightness up or down, without accounting for the relatively faint luminosity observed specifically in `Inner Arm'. If the SFH of M31 derived from resolved stellar color-magnitude diagrams (CMD) in \citealt{Williams2017PHAT} is reliable, the dimness of `Inner Arm' may suggest that its LMXB population is fainter than expected for populations {older than $\sim$1 Gyr} in Fig.~\ref{pic:SFH}a. 

To investigate and illustrate this further, we develop a toy model of luminosity evolution without considering the metallicity. We design our luminosity evolution model with 6 free parameters indicating the luminosity at specific distributed ages between 0.003 and 10 Gyr. We generate predictions by convolving the luminosity evolution models with the SFH specific to each of our four defined regions. For each region, the SFH employed is the average of the results from the four different stellar evolution models. These predictions are then compared with our measurements. Although the model parameters are degenerate since we only have 4 measurements, we can still utilize MCMC methods to constrain the relative trend of the luminosity evolution. The resulting model is presented in Fig.~\ref{pic:Luminosity_model}. Our model shows a higher luminosity than \cite{Fragos2013} in the HMXB-dominated era (<150 Myr). From this, we estimate the HMXB contribution to the total luminosity to be $\sim$5\%, a fraction that remains minimal, assuming a constant LMXB output for the first 500 Myr. The key finding from this comparison, however, is a more pronounced dimming of LMXB luminosity approximately 1 Gyr after a star formation burst compared to the widely used \cite{Fragos2013} model (Fig.~\ref{pic:Luminosity_model}, black solid line). Notably, this faster decline in LMXB luminosity around ~1 Gyr is consistent with the independent findings of \cite{Gilbertson2022} (Fig.~\ref{pic:Luminosity_model}, green data points with error bars), lending further support to our result.

This finding provides a unified physical explanation for two key observations. Firstly, it directly accounts for the observed faintness of the `Inner Arm' region. The distinct star formation history of the `Inner Arm', which is heavily weighted towards the 2–4 Gyr age range, means its X-ray luminosity is particularly sensitive to the rapid fading our model predicts after 1 Gyr. Secondly, on a galactic scale, this same insight explains why M31 as a whole exhibits a systematically lower $\alpha_\mathrm{LMXB}$ compared to other galaxies. Since the bulk of M31's total stellar mass was also formed more than 1 Gyr ago, the diminished contribution from this vast, older stellar population naturally leads to a lower integrated X-ray luminosity. Thus, our new luminosity evolution model, constrained by the diverse SFHs within M31, provides a self-consistent reason for both the relative faintness of the `Inner Arm` and the under-luminous nature of the entire galaxy.

The rapid decline in LMXB luminosity around 1~Gyr revealed by our toy model is highly consistent with recent findings from large galaxy surveys. Both \citet{Gilbertson2022} and \citet{Lehmer2024} independently observed that the $L_\mathrm{X}/M_*$ at stellar population ages of 0.3--3~Gyr is significantly lower than predicted by the theoretical models of \citet{Fragos2013}. They attribute this discrepancy to an overestimation in the population synthesis prescriptions for LMXBs originating from intermediate-mass donor stars (approx. 2--8~$M_{\odot}$). \citet{Gilbertson2022} note that this particular evolutionary stage, where intermediate-mass stars initiate a short-lived, high-accretion phase, is especially difficult to model accurately. Our study, through the detailed analysis of M31's sub-regions, provides new, independent evidence supporting this conclusion and highlights the necessity of recalibrating key physical processes in binary evolution models, for which our work provides a new observational benchmark.

\begin{figure}
    \centering
    \includegraphics[width=0.95 \linewidth]{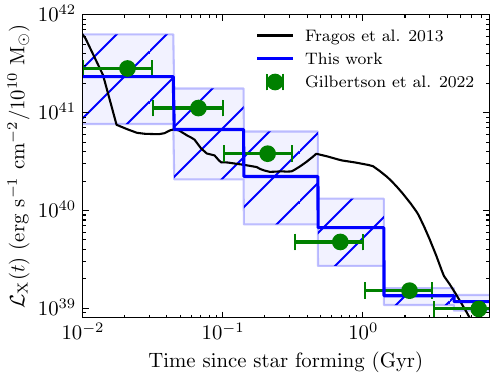}
    \caption{{Luminosity evolution model of X-ray binaries, presented in terms of bolometric luminosity. The black solid line shows the theoretical prediction from \citealt{Fragos2013}. Our new model, constrained in this work, is represented by the blue shaded region, which denotes the 68\% confidence interval of the median luminosity evolution. The green data points show observational constraints from  \cite{Gilbertson2022}) for stellar populations with known ages. Compared to the \citealt{Fragos2013} model, our work predicts a systematically lower luminosity at $\sim$1 Gyr, suggesting a more rapid decline of XRB luminosity which is in good agreement with the trend from \citealt{Gilbertson2022}}}
    \label{pic:Luminosity_model}
\end{figure}

\section{Summary}

In this study, we analyze the X-ray luminosity function (XLF) of 4,506 X-ray sources around M31 as part of the New-ANGELS legacy survey, providing one of the most detailed population studies of X-ray sources in M31 to date. Using data from various regions within M31 (including the central, disk, and halo regions) we construct incompleteness-corrected XLFs, allowing for the separation of source populations and comparison across distinct regions. The northern and southern regions serve as valuable baselines for background and foreground source modeling.

Our findings reveal that M31's X-ray source population is dominated by low-mass X-ray binaries (LMXBs), with a minimal contribution from high-mass X-ray binaries (HMXBs). We demonstrate that the galaxy's systematically low $\alpha_\mathrm{LMXB}$ can be explained by a more rapid fading of LMXB populations than previously modeled. Our analysis, constrained by the differing star formation histories of M31’s sub-regions, indicates a pronounced dimming for stellar populations aged $\sim1$ Gyr. This faster decline not only explains the relative faintness of the `Inner Arm` region, which has a significant 2-4 Gyr stellar population, but also accounts for the under-luminous nature of the galaxy as a whole, since the majority of its total stellar mass formed in this epoch or earlier.

Notably, the differences in $\alpha_\mathrm{LMXB}$ and the XLF shape across M31 regions suggest that the `universal' correlation between integrated luminosity and stellar mass shows considerable scatter, even within different regions of the same galaxy. The lower $\alpha_\mathrm{LMXB}$ of `Inner Arm' further implies that the LMXBs should be much fainter than what is expected from our preliminary luminosity evolution model, especially those  $~1$ Gyr. 

\section*{Acknowledgements}

We would like to thank Chengzhe Li, Junjie Mao, Marat Gilfanov and Zhongli Zhang for helpful discussions, and Thomas H. Jarrett for providing high-quality \emph{WISE} image of M31. RH acknowledges support from the China Scholarship Council. 
This work is supported in part by the China Manned Space Program through grant no. CMS-CSST-2025-A10, by the Ministry of Science and Technology of China through grant no. 2018YFA0404502, and by the National Natural Science Foundation of China through grant no. 11821303. J.T.L. acknowledges the financial support from the National Science Foundation of China (NSFC) through grants 12273111 and 12321003, and also the China Manned Space Program through grant no. CMS-CSST-2025-A04.

\section*{Data Availability}
The X-ray data that forms the basis of this article can be accessed in the \emph{XMM-Newton} data archive, which is available at \url{https://nxsa.esac.esa.int/nxsa-web/}.
The catalogue of X-ray sources around M31 is openly accessible in \citetalias{huangrui2023APJS}.
The \emph{WISE} data can be obtained from SkyView, available at \url{https://skyview.gsfc.nasa.gov/}. The code and data (including the sensitivity map) used for the analysis in this paper are available at \url{https://github.com/RuiHuangAstro/New-ANGELS_II}.

\bibliographystyle{mnras}
\bibliography{M31}

\appendix

\section{Sensitivity map}
\label{sec:sensitivity_map}
The calculation of the sensitivity is the reverse process of the source detection, so we first introduce the details of the source detection. 
\texttt{edetect\_stack} calls \texttt{emldetect} to perform the maximum likelihood PSF fitting over the assumed source position. To reach the best-fitting source position and count rate (and extent radius for extended source), \texttt{emldetect} minimises the C-statistic (\citealt{cashParameterEstimationAstronomy1979})
\begin{equation}
C=2\sum_{i=1}^{N}(e_{{i}}-n_{i}~ln~e_{i})
\label{eq.C_stat}
\end{equation}
where $n_{i}$ and $e_{i}$ are the measured counts and the expected counts in pixel $i$ respectively, $N$ is the number of pixels selected for fitting, which is determined by \texttt{eml\_ecut} of \texttt{edetect\_stack} (cut-off radius in pixels used for calculation, \texttt{eml\_ecut}=15 in this work).
For a source in image coordinate (x,y), the expected counts $e_{i}$ in pixel $i$ is calculated as 
\begin{equation}
e_{i} = rate(x,y) \times PSF_{i}(x,y) \cdot exp_{i} + bkg_{i}
\label{eq.expected_counts}
\end{equation}
where $rate(x,y)$ is the count rate of the source, $PSF_i(x,y)$ is the probability of collecting photons in pixel $i$ for the point source at the position (x,y) according to the PSF model, $exp_{i}$ and $bkg_{i}$ are the effective exposure time and the background counts in pixel $i$. 

According to the theorem of Wilks, the difference between the C-statistic 
\begin{equation}
\Delta C= C_{\mathrm{null}}-C_{\mathrm{best}}
\label{eq.delta_C}
\end{equation}
follows $\chi^2$ distribution with degree of freedom $\nu$, where $C_{\mathrm{best}}$ is the minimized $C$, \begin{equation}
C_{\mathrm{null}}=2\sum_{i=1}^{N}(e_{i}-b_{i}~\mathrm{ln}~e_{i})
\label{eq.C_null}
\end{equation}
is C-statistic for null hypothesis model that the signal is caused by background fluctuation $b_{i}$.  Then the probability of obtaining the detected counts by background fluctuation is defined as 
\begin{equation}
P(\chi^2 \geq \Delta C)=1-P_{\Gamma}(\frac{\nu}{2},\frac{\Delta C}{2}) 
\label{eq.P}
\end{equation}
where $P_{\Gamma}$ is the regularized lower incomplete gamma function, $\nu=n+2$ is the free parameters used during fitting point source ($n$ parameters refers to the $n$ count rates from $n$ images and 2 parameters for position). The likelihood is defined as 
\begin{equation}
L=-ln(P)
\label{eq.L}
\end{equation} 

In order to calculate the sensitivity map at the requested likelihood $L$, we can derive the expected $\Delta C$ based on Eq.~\ref{eq.P} and \ref{eq.L}. Then at the condition of the given background $b_i$, we calculate Eq.~\ref{eq.delta_C}  with Eq.~\ref{eq.C_null} and 
\begin{equation}
C_{\mathrm{best}}=2\sum_{i=1}^{N}(e_{i}-e_{i}~ln~e_{i}) 
\label{eq.C_best}
\end{equation} 
where we assume a perfect fitting with $n_i=e_i$ in Eq.~\ref{eq.C_stat}.
Based on the Eq.~\ref{eq.expected_counts} to \ref{eq.C_best}, we have the function of the likelihood on the expected count $e_i$. 
So the sensitivity of the given likelihood at given region is produced.

The background for discrete sources includes particle background, sky background, and any diffuse emission around the sources. We derive the background map used for calculating the likelihood of the detected sources using \texttt{esplinemap}, which generates spline background maps from non-source regions.
After iterating through every pixel, we generate the sensitivity map. It's important to note that $\nu=n+2$ adapts dynamically based on the number of images (e.g., $n=6$ for three instruments from two observations) used in a particular position. This adaptability ensures the applicability of the algorithm not only for single observations but also for multiple overlapping observations.

\section{Simulation of source detection}
\label{sec:simulation}

In order to examine our sensitivity map method, 
we perform source detection simulation by overlaying fake sources over three real overlapping observation (0800732101, 0800732201, and 0800732501) which are part of our observations.

First, we generate a fake point source list that contains the position and the X-ray flux in each band. The number of sources in the list is limited to be within 10\% of the existing source number in the observations, so that the field does not become too crowded for source detection. These fake sources have uniform spatial distribution and uniform X-ray flux distribution in logarithmic scale (power law distribution with index as -1). 
Second, we generate the source model image. We produce point spread function(PSF) model image(Fig.~\ref{pic:Simulation}a) with task \texttt{psfgen} for the source at the dedicated position, then multiply it with the designed count rate (convert from the X-ray flux in fake sources list) and exposure map (b in Fig.~\ref{pic:Simulation}) at the same position. 
Third, we poissonize the source model image (Fig.~\ref{pic:Simulation}b) to source count image  (Fig.~\ref{pic:Simulation}c). 
Fourth, we overlay the simulated source count image over the original count map. By overlaying the fake sources on the original images (Fig.~\ref{pic:Simulation}d), we have the identical background map as in the observation. Finally, we repeat the above procedures for all fake sources in $n\times3\times5$ count maps (n observations, 3 instruments, and 5 bands). 
After we generate the count images, we perform the source detection via task \texttt{edetect\_stack} as described in \citealt{huangrui2023APJS} 1000 times. 

\begin{figure*}
\begin{center}
\includegraphics[width=1.0 \columnwidth]{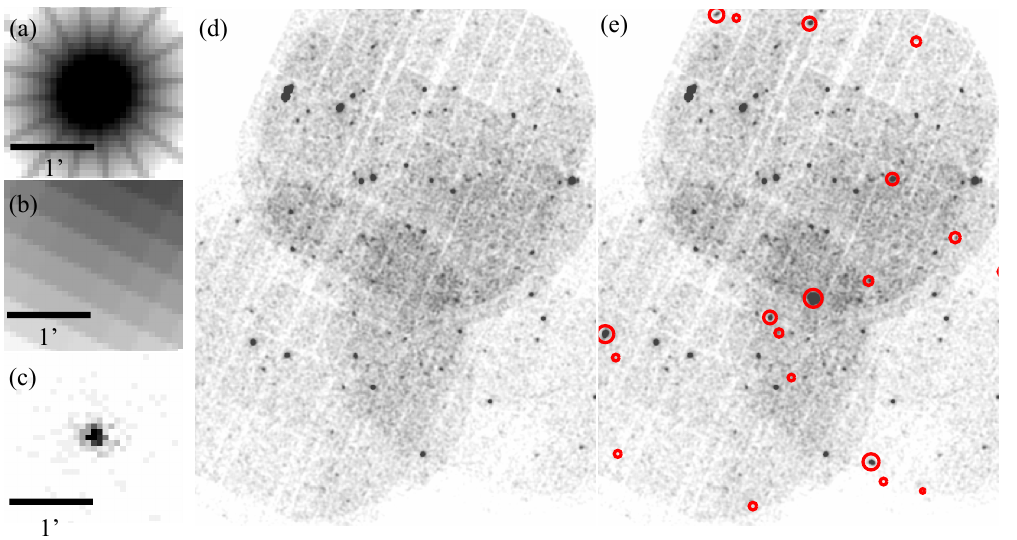}
\end{center}
\caption{a: PSF model generated with \texttt{psfgen}. b: exposure map. c. image of a faked source ($\mathrm{count~s^{-1}}$).  d: mosaiced EPIC count image of 3 observations with overlapping. e: the same d but overlaid with fake sources. The red circles denotes the positions of fake sources.} \label{pic:Simulation}
\end{figure*}

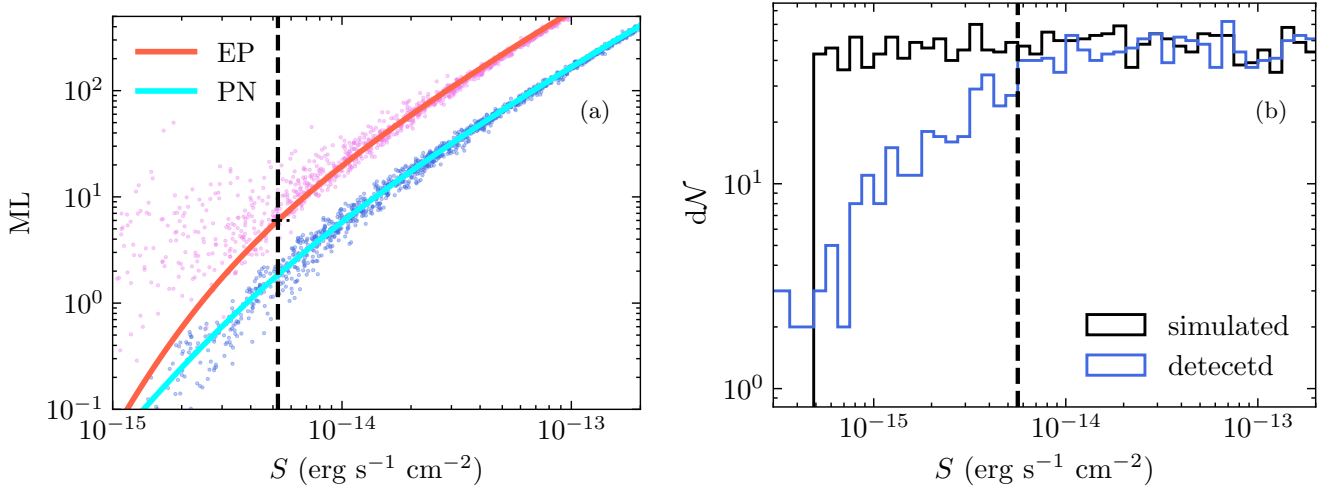
\begin{figure*}
    \centering
    \includegraphics[width=0.48\linewidth]{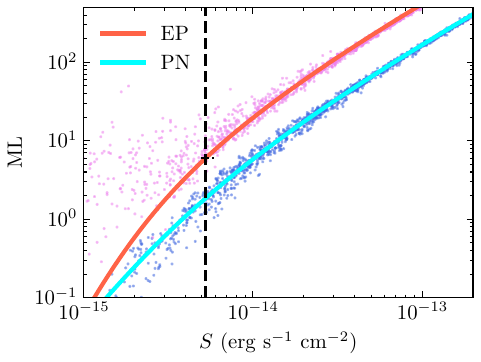}
    \begin{picture}(0,0)
    \put(-30,145){\makebox(0,0)[l]{{\color{black}(a)}}}
    \end{picture}
    \includegraphics[width=0.48\linewidth]{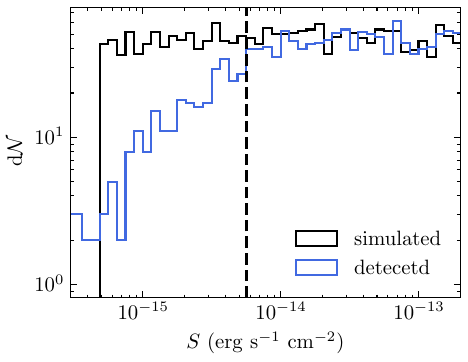}
    \begin{picture}(0,0)
    \put(-30,145){\makebox(0,0)[l]{{\color{black}(b)}}}
    \end{picture}
    \caption{Left: The maximum likelihood (ML) and the detected X-ray flux for the simulated sources at the the same position. The solid line represent the predicted relation between the likelihood and the detected X-ray flux for EPIC(red) and PN (cyan). EPIC means the detection is based on both mos1, mos2, and pn observations. Right: the true X-ray flux distribution (black) and the detected EPIC X-ray flux distribution (blue) of simulated sources. The black dashed lines in two panels refer to the sensitivity at EPIC which we used to filter the sources. The sensitivity is obtained assuming likelihood equals 6.}
    \label{fig:ML_S_ScatterPlot}
\end{figure*}

Based on the equation in Appendix.~\ref{sec:sensitivity_map}, we can predict the maximum likelihood (ML)-flux relation for any position.
In Fig.~\ref{fig:ML_S_ScatterPlot}a, we provide an example that our predicted relation closely follows the detection results for one position (see Fig.~\ref{fig:ML_S_ScatterPlot}a). In Fig.\ref{fig:ML_S_ScatterPlot}b, the detection probability at the position drops near a certain sensitivity threshold. The sensitivity can be obtained from ML-flux relation in Fig.~\ref{fig:ML_S_ScatterPlot}a at different ML assumptions. This indicates that not all simulated sources are detectable with 100\% certainty below a specific X-ray flux. Our empirically chosen ML value of 6 (black dashed line in Fig.~\ref{fig:ML_S_ScatterPlot}) ensures that sources above the sensitivity threshold are detectable, while a lower X-ray flux limit may not be appropriate.

\begin{figure}
\centering
\includegraphics[width=.6\textwidth]{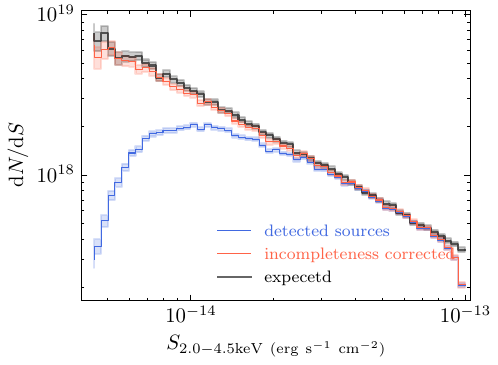}
\caption{The X-ray flux distributions of the simulated sources. The X-ray flux distribution of detected fake sources is shown in blue line. The corresponding incompleteness corrected X-ray flux distribution is in red line. 
The expected X-ray flux distribution of these simulated sources is in black.}
\label{pic:simulation}
\end{figure}

In Fig.~\ref{pic:simulation}, we show the X-ray flux distribution of the all simulated sources. The recovered X-ray source flux distribution closely matches the expected distribution. However, there is a slight discrepancy caused by the 'zero flux issue' in source detection. To be specific, for some faint sources, the X-ray flux detected by one of the instruments is recorded as zero. The combined X-ray flux is then calculated using the detections from three instruments (MOS1, MOS2, and pn), resulting in the combined X-ray flux being 2/3 of the average X-ray flux from the other two instruments. If we exclude these simulated sources with 'zero flux', the recovered X-ray flux distribution aligns very well with the expected distribution.
Our sensitivity map method can not resolve this detection issue. However, according to our simulations, the biased slope resulting from this issue is within 0.05, which is much smaller than the uncertainty in our fitting. Therefore, we choose to disregard this issue in our analysis.

Besides the issue above, our method does not consider the source confusion problem and the detection uncertainty.

\section{Stellar mass and star formation rate}
\label{sec:stellarmass_SFR}

The stellar mass and SFR are derived on \emph{WISE} image data \citep{Jarrett2013,Jarrett2019,Vargas2018,Wang2021}. The formula for calculating stellar mass:
\begin{equation}
\mathrm{log}M_{\mathrm{\star}}/L_{\mathrm{W1}}=-2.54(M_\mathrm{W1}-M_\mathrm{W2})-0.17
\label{eq:stellar_mass}
\end{equation}
Here, $L_{\mathrm{W1}} (\mathrm{L}_{\odot,\mathrm{W1}}) = 10^{-0.4(M_\mathrm{W1}-\mathrm{M_{\odot,W1}})}$, where $M_\mathrm{W1}$ and $M_\mathrm{W2}$ are the absolute magnitude of \emph{WISE} W1($3.4~\mathrm{\mu m}$ ), W2($4.6~\mathrm{\mu m}$ ) bands and $\rm M_{\odot,W1} = 3.24$ represents the solar value in the W1 band. The distance modulus we adopted here is 24.4.
The estimation of the SFR is calculated as follows:
\begin{equation}
    SFR(\mathrm{M_{\odot}~yr^{-1}}) = 7.5 \times 10^{-10}L_\mathrm{W4}(\mathrm{L}_{\odot})
\end{equation}
Where the $L_\mathrm{W4}$ is the luminosity in W4 ($22~\mathrm{\mu m}$) band and solar luminosity is denoted as \(\mathrm{L}_{\odot} = 3.839 \times 10^{33}~\mathrm{erg~s^{-1}}\).

The luminosity \(L_\mathrm{W4}\) (in \(\mathrm{L}_{\odot}~\mathrm{arcsec}^{-2}\)) is related to the specific intensity \(\nu I_\mathrm{W4}\) (where \(\nu=1.3 \times 10^{13} \mathrm{~Hz}\) is the central frequency corresponding to the \(22.8\rm~\mu m\) wavelength of the \emph{WISE} W4 band; \citealt{Brown2014}) through the equation:

\begin{equation} \label{eq1}
\begin{split}
L_\mathrm{W4}  & =\nu \times 5.2 \times 10^{-5} I_\mathrm{W4} (4 \pi D^2) /(10^{23} \times 3.8 \times 10^{33}) \\
        & =4.4 \times 10^3 I_\mathrm{W4},
\end{split}
\end{equation}

Here, the term \(5.2 \times 10^{-5}\) converts \(\rm DN~s^{-1}\) to \(\rm Jy\) \footnote{\url{https://wise2.ipac.caltech.edu/docs/release/allsky/expsup}} , \(10^{23}\) converts \(\rm Jy\) to \(\rm erg~s^{-1}~cm^{-2}~Hz^{-1}\), and the distance D to M31 is assumed to be 761 kpc. Combining these equations, we derive 
\begin{equation}
    \sum_{\mathrm{SFR}}(\mathrm{M_{\odot}~yr^{-1}arcsec^{-2}}) = 3.3 \times 10^{-6}I_{22}~(\mathrm{DN s^{-1} arcsec^{-2}})
\end{equation}
This formula enables the calculation of the star formation rate based on the measured integrated intensity (\(I_\mathrm{W4}\)) in the image.

\begin{figure*}
    \centering
    \includegraphics[width=0.31 \linewidth]{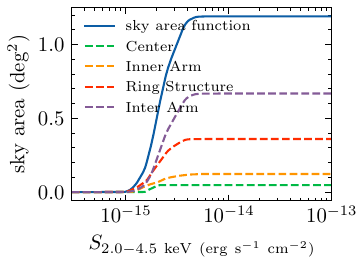}
    \begin{picture}(0,0)
    \put(-132,40){\makebox(0,0)[l]{{\color{black}(a)}}}
    \end{picture}
    \includegraphics[width=0.31 \linewidth]{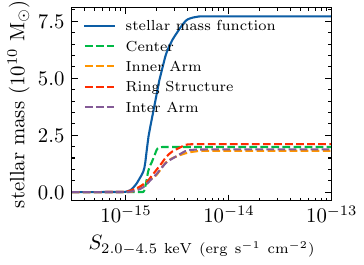}
    \begin{picture}(0,0)
    \put(-130,40){\makebox(0,0)[l]{{\color{black}(b)}}}
    \end{picture}
    \includegraphics[width=0.31 \linewidth]{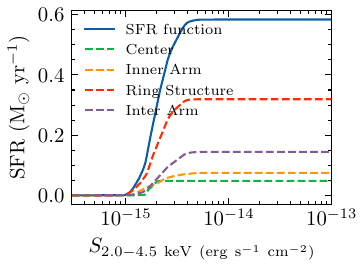}
    \begin{picture}(0,0)
    \put(-130,40){\makebox(0,0)[l]{{\color{black}(c)}}}
    \end{picture}
    \caption{The sky coverage function, stellar mass function, and SFR function in different regions. The solid line and dashed lines represent the combined and individual functions in `Center', `Inner Arm', `Ring Structure', and `Inter Arm'.}
    \label{pic:stellar_mass_and_SFR}
\end{figure*}
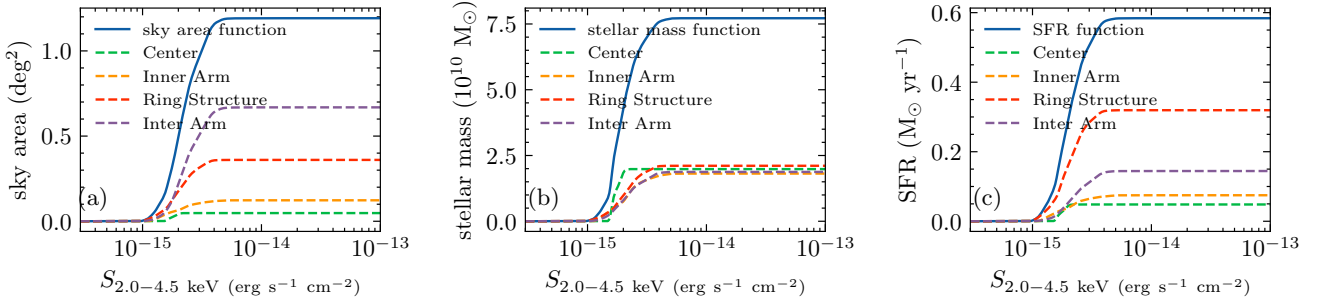

\end{document}